\documentclass[11pt, final]{myarticle} 
\usepackage{setspace}
\usepackage{amssymb,cite,latexsym,amsmath,subfigure}
\usepackage{graphicx}
\usepackage{psfrag}
\usepackage{pstricks}
\usepackage{wrapfig}
\makeatletter
\renewcommand\@seccntformat[1]{\csname the#1\endcsname.\quad}

\date{}
\begin{document} 

\title{Stochastic rocket dynamics under random
nozzle side loads: Ornstein-Uhlenbeck boundary layer
separation and its coarse grained connection to
side loading and rocket response}

\author{R. G. Keanini\footnote{ Professor, Mechanical Engineering, 
9201 University City Blvd., DCH 353; rkeanini@uncc.edu}\\
Nilabh Srivastava\footnote{ Assistant Professor, Mechanical Engineering, 9201 
University City Blvd., DCH 355; nsrivast@uncc.edu} \\
Peter T. Tkacik\footnote{ Assistant Professor, Mechanical 
Engineering, 9201 University 
City Blvd., DCH 354; ptkacik@uncc.edu} \\
Department of Mechanical Engineering and Engineering Science\\
The University of North Carolina at Charlotte \\
Charlotte, North Carolina 28223-0001 \\
David C. Weggel\footnote{ Associate Professor, Civil
Engineering, 9201 University 
City Blvd., Cameron 144; dcweggel@uncc.edu} \\
Department of Civil and Environmental Engineering \\
The University of North Carolina at Charlotte \\
Charlotte, North Carolina 28223-0001 \\
P. Douglas Knight\footnote{Director, 
500 West Broad Street, knight@mitchell.cc.nc.us}\\
Mitchell Aerospace and Engineering \\
Statesville, North Carolina 28677}

\maketitle

\newpage

\begin{abstract}
A long-standing, though ill-understood
problem in rocket dynamics, rocket 
response to random, altitude-dependent nozzle side-loads, is investigated.
Side loads arise during low 
altitude flight due to random, asymmetric, shock-induced 
separation of in-nozzle boundary layers.
In this paper, stochastic evolution of the in-nozzle boundary layer
separation line, an essential feature underlying
side load generation, is connected to random, altitude-dependent
rotational and translational rocket response via a set
of simple analytical models. 
Separation line motion, extant on a fast boundary
layer time scale,
is modeled as an 
Ornstein-Uhlenbeck process.
Pitch and yaw responses, taking place
on a long, rocket dynamics time scale,
are shown to likewise evolve as OU processes.
Stochastic, altitude-dependent rocket translational motion
follows from linear, asymptotic versions of the
full nonlinear equations of motion; the model is
valid in the practical limit
where random pitch, yaw, and roll rates all
remain small.
Computed altitude-dependent rotational and translational velocity
and displacement statistics are compared against
those obtained using recently reported high fidelity simulations
[Srivastava, Tkacik, and Keanini, \textit{J. Applied Phys.}, \textbf{108,}
044911 (2010)];
in every case, reasonable agreement is observed.
As an important prelude,
evidence indicating the physical consistency 
of the model
introduced in the above article is first presented: it is shown that
the study's
separation line model allows
direct \textit{derivation} of experimentally observed
side load amplitude and direction densities.
Finally, it is found that the analytical models proposed in this paper
allow straightforward identification
of practical approaches for: i) reducing pitch/yaw response
to side loads, and ii) enhancing pitch/yaw damping once side loads
cease.
\end{abstract}

\vspace{.5cm}


\section{Introduction}
Although the recorded history of powered rocket flight spans a millenium, 
dating to eleventh century China \cite{spacecraft-early-years}, and 
experimentally-based rocket development traces two hundred years to 
eighteenth century England \cite{spacecraft-early-years}, and while an 
enormous scientific and engineering literature attaches to the dynamics, 
design, and control of rockets, missiles, and spacecraft, numerous 
interesting, practically important questions remain.
One of the most basic and well-studied concerns 
prediction of rocket trajectories during ascent.  While the 
long time success of large rocket and missile programs suggests that 
this essential question has been solved, in fact,
a variety of difficult-to-predict features introduce
significant uncertainty. These
include: i) altitude-, attitude-, 
and speed-dependent aerodynamic forces \cite{rocket-propulsion-elements}, 
ii) random loads produced by location- and altitude-dependent 
wind and atmospheric 
turbulence \cite{gust-model-nasa, wind-data-1988, wind-data-1999,
nasa-atm-turbulence-model}, 
iii) rocket design and construction imperfections
\cite{ref18}, iv) fuel sloshing (liquid 
fuel rockets) \cite{slosh-1959,slosh-1967}, v) slag formation (solid fuel 
rockets) \cite{rocket-propulsion-elements}, and vi) random impacts with 
air-borne animals and debris \cite{animal-impact}.

Random, nozzle side loads present
a further, singularly ill-understood feature complicating
ascent prediction. 
As depicted in Fig. 1, and as 
described, e.g., in \cite{nilabh2010, keanini},
nozzle side loads
appear in over-expanded nozzles during low-altitude flight, when
high ambient pressure forces external air upstream into the 
nozzle. Under these conditions,
the inflow overcomes the low pressure, low inertia, 
near-wall nozzle out-flow. At a locus of points,
the nominal, instantaneous boundary layer separation line,
inflow inertia decays sufficiently that outflow inertia
can turn the inflow back on itself; the reversed
inflow forms a virtual corner and, in turn,
an oblique, circumferential shock. See Fig. 2.  

\renewcommand{\thefigure}{\arabic{figure}}
\setcounter{figure}{0}
\begin{figure}[!hb]
\centering
\includegraphics[width=4.75in]{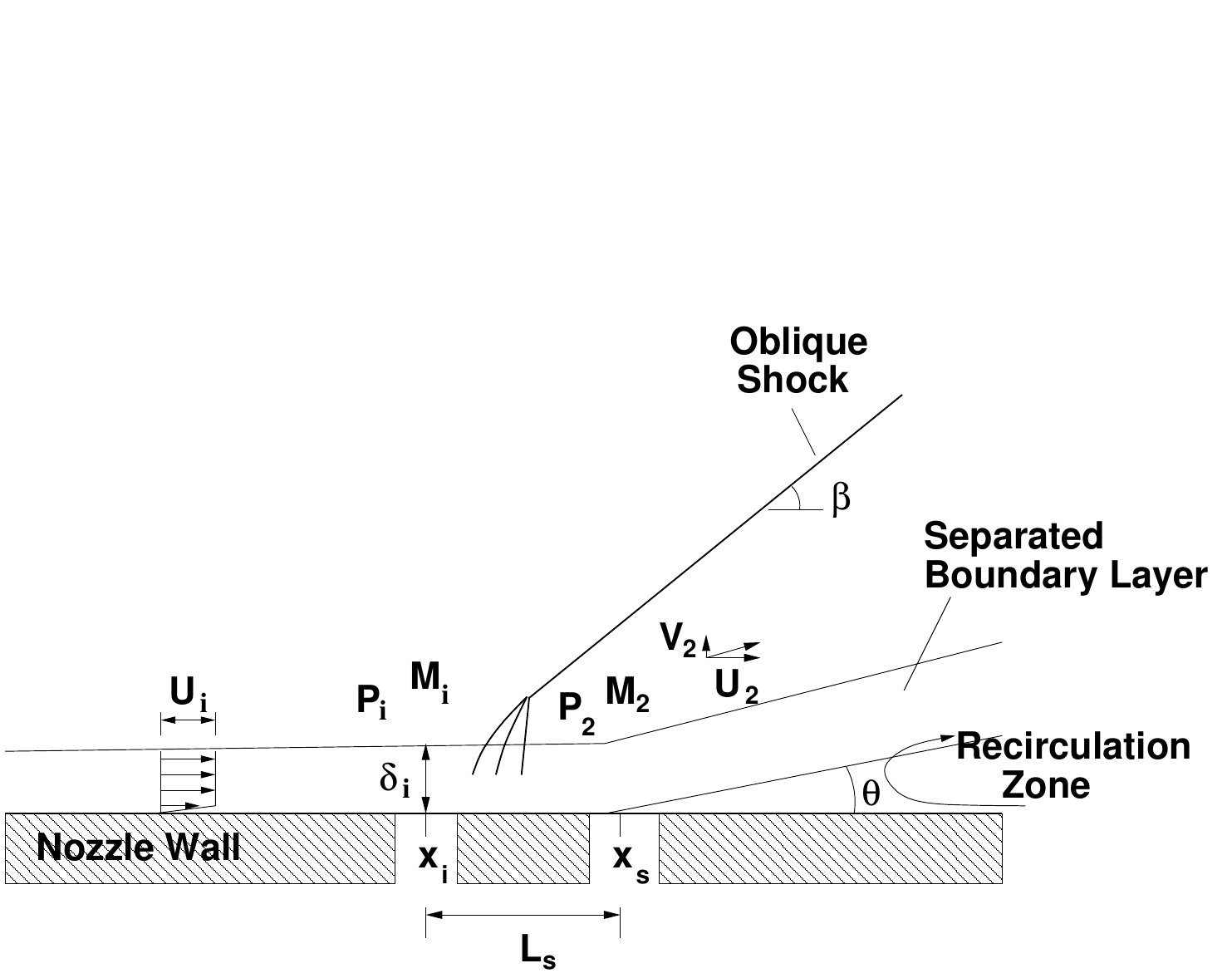}
\caption{Shock-induced boundary layer separation
in overexpanded supersonic nozzle flow. The process 
typically occurs during low altitude flight when 
ambient pressure is high enough to force
atmospheric air into the nozzle. The incoming air
flows upstream along the low-inertia, near-wall region until
downstream-directed
boundary layer inertia turns it, forming a virtual
compression corner. An
oblique shock thus forms, and the combined action of
shock-induced pressure rise and inertial pressurization
produced by the inflow forces 
the down-flow boundary layer to separate. Pressures, mach numbers,
and velocities are 
denoted, respectively, by $ P , $ $ M , $ and $ U $ and $ V . $
Axial positions where the boundary layer
starts to thicken (\textit{i} denotes \textit{incipient}),
and where it separates are denoted, respectively,
as $ x_i $ and $ x_s ; $ the nominal shock-boundary layer
interaction zone is shown as $ L_s . $  
Since the separation line position, $ x_s , $
and downstream conditions
vary with the altitude-dependent ambient pressure, $ P_a = P_a (H(t)) $ 
\cite{keanini}, all variables shown likewise
vary with $ H(t) . $ }
\end{figure}

Random side loads appear due to the pressure jump,
$ \Delta P , $ extant across the asymmetric, 
stochastically evolving separation line.
Viewing the
instantaneous separation line as 
the superposition of the line's slowly moving
mean axial position,
$ x_s (t) , $ and it's rapid random fluctuation about $x_s(t) , $
one observes that since $ x_s (t) $ and $ \Delta P $
are both altitude-dependent, random side loads, 
and resulting torques and
translational and rotational rocket responses,
are all likewise altitude-dependent.

\begin{figure}[!hb]
\centering
\includegraphics[width=5.5in]{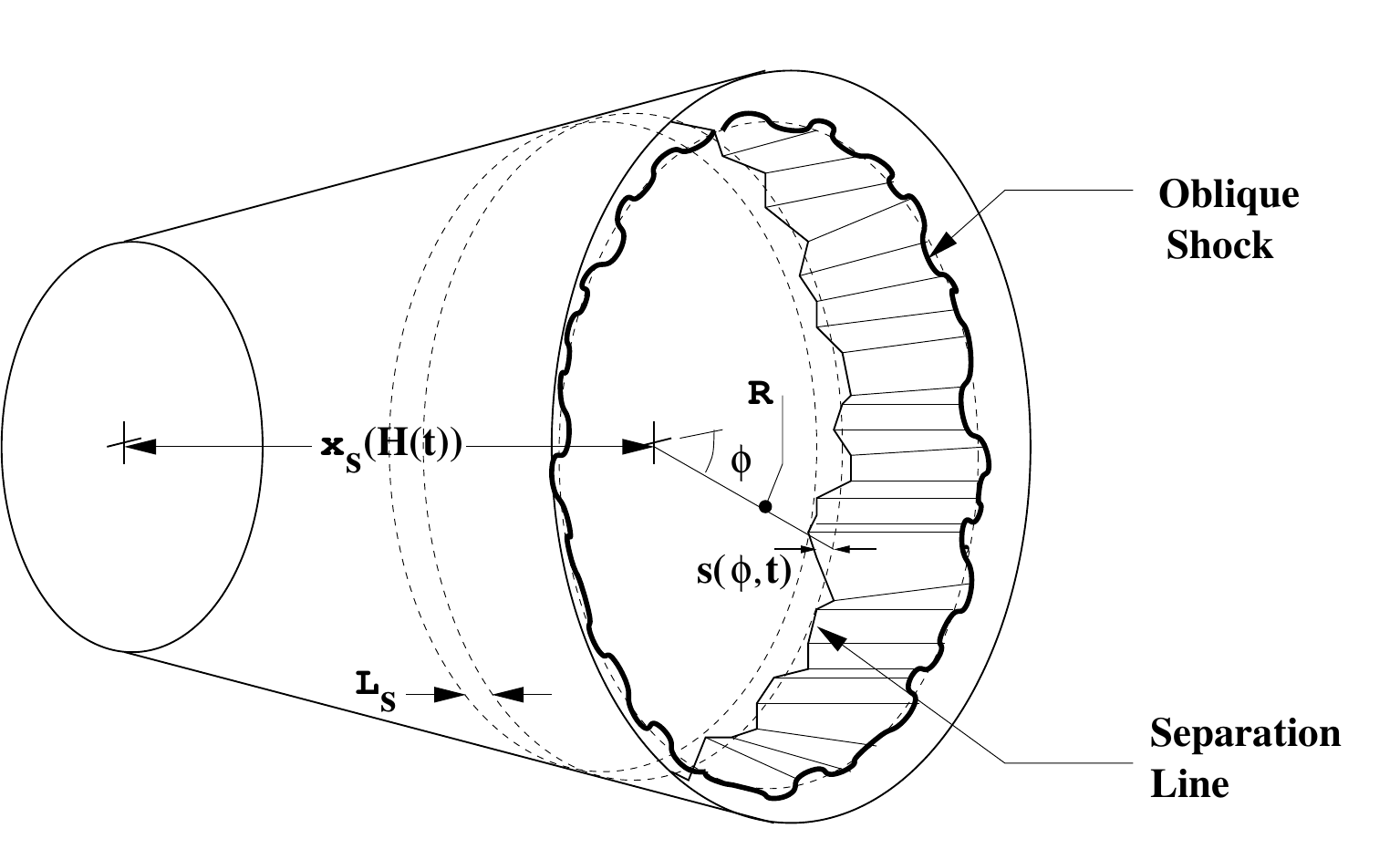}
\caption{Schematic of stochastic boundary layer separation line
and associated, rippled, azimuthal oblique shock. 
The mean separation line position relative to the nozzle throat,
$ x_s , $ varies with rocket altitude, $ H(t) ; $ the corresponding
nozzle radius is $ R = R(H(t)) . $
The instantaneous separation
line position relative to $ x_s (t) $
is shown as $ s(\phi, t) .$ The separation line 
lies on the nozzle wall and, in a nominally symmetric nozzle,
the shock forms an azimuthally independent, average angle
which varies with $ x_s (t) . $}
\end{figure}

Supersonic flow separation and side load 
phenomena in nozzles, studied under fixed external pressure, i.e.,
effectively fixed altitude conditions,
has attracted significant 
attention \cite{ref1, ref2, ref3, ref4, ref6, ref7, ref7a, ref7b, ref8, ref16};
Ostlund and Muhammad-Klingmann \cite{ref6a} review much of this work.
Numerical and experimental studies
have been reported, for example, by Frey and Hagemann \cite{ref3, ref16a},
Onofri and Nasuti \cite{ref16b}, 
Pekkari \cite{ref16c}, Schwane and Xia \cite{ref16d}, and Shimizu et al.
\cite{ref16e}.
Fundamental 
studies of shock-boundary layer 
interactions \cite{ref9, ref10, ref11, ref12, ref14, ref17}
in which (typically oblique) shocks are 
produced by a variety of shock generators, 
have, in turn, provided essential insight
into the fluid dynamic  features underlying shock-induced separation.

Nozzle side loads can be perilous, damaging nozzles and/or 
attached equipment, or inducing catastrophic 
excitations within structural components; the failure of the 
Japanese H-II in 1999, for example, has been attributed to
nozzle side loads \cite{sekita}. In large engines, side 
load magnitudes can be extreme, on the order of 
250,000 pounds, for example, in Saturn V engines \cite{ref18}.
Thus, minimizing and designing to 
accommodate side loading represents an essential
rocket design task.  

While side loads have long complicated ascent prediction,
until recently \cite{nilabh2010}, their effects on
rocket dynamics have remained completely unknown.
The work in \cite{nilabh2010} focused on the 
stochastic ascent of sounding-rocket-scale rockets
subject to altitude-dependent random side loads and 
torques. A high fidelity numerical model was developed which 
incorporated the following features:
\renewcommand{\labelenumi}{\alph{enumi})}
\begin{enumerate}
\item a variable mass, six-degree-of-freedom, nonlinear
rocket dynamics model \cite{ref19}, 
\item empirical models of the altitude-dependent ambient
atmosphere \cite{wind-data-1999} and mach number-dependent
drag coefficient \cite{ref20},
\item a model of separation line motion which
incorporated experimentally observed
statistical properties of
shock-separated flat plate turbulent boundary layers
\cite{ref11, ref12, ref17}, and
\item an \textit{ad hoc,} though physically consistent model
of side load statistics. 
\end{enumerate}
The model allowed Monte Carlo simulation of random rocket ascents,
as well as estimates of altitude-dependent, ensemble-averaged
translational and rotational velocities and displacements. 
\subsection{Objectives}
Through the remainder of the paper, we will
often refer collectively to the set of numerical models described in
\cite{nilabh2010} as \textit{Model I}; the present
set of analytical models will likewise be referred to as 
\textit{Model II.}
\renewcommand{\labelenumi}{\Alph{enumi})}
\begin{enumerate}
\item The first, and most crucial objective
centers on establishing the \textit{physical consistency,} i.e., the congruence
with physical principles and available experimental data,
of both Models I and II.
Due to the difficulties associated with 
detecting and isolating
dynamic side loads within nozzles subject to simultaneous
random aerodynamic and structural loading,
obtaining experimental data on the effects of side loading
on ascent will likely 
remain problematic.
Indeed, no such data presently exists.

\vspace{.35cm} 

Thus, three key features motivate pursuit of this objective: \\
i) substantial experimental challenges will likely continue
to limit direct measurements, \\
ii) theoretical insight is presently limited to the numerical models
comprising Model I \cite{nilabh2010}, and crucially,\\ 
iii) demonstration of physical consistency establishes a 
reasonable foundation for further work on this
long-standing problem.

\vspace{.35cm} 

The first objective is pursued in
two steps. 
\begin{enumerate}
\item We first show
that the separation line model introduced in \cite{nilabh2010}
allows \textit{derivation} of both the 
\textit{assumed} side load model in \cite{nilabh2010},
as well as 
experimentally observed side load
amplitude and direction densities \cite{ref7, ref7a, ref7b}.  
See section 2.

\vspace{.3cm} 

Since Model I \cite{nilabh2010} incorporates a high fidelity
rocket dynamics model and reasonable models
of rocket aerodynamics and altitude-dependent ambient
atmospheric conditions, 
and since the above demonstration
ties the most
uncertain portions 
of Model I, i.e.,
the separation line and side load models,
to experimental observations \cite{ref7, ref7a, ref7b},
we argue that Model I \cite{nilabh2010} 
thus represents a physically consistent description
of rocket ascent under altitude-dependent side loads.

\item Presuming the physical consistency of Model I \cite{nilabh2010},
we then argue that since Model II 
predicts altitude-dependent 
rotational and translational rocket-response
statistics that are, in every instance examined, 
consistent with those estimated via Model I (see section 7), 
Model II is likewise consistent.
\end{enumerate}

\item It is found that the simple models comprising
Model II provide a straightforward framework
for interpreting observed random rocket responses
to side loading, as well as for identifying 
practical approaches for designing against 
side loads. See section 7.
The second objective thus centers on highlighting and exploiting
this theoretical framework.
\end{enumerate}

\subsection{Overview} 
The following models and interconnecting elements tie
stochastic separation line motion to rocket response, and thus
comprise the paper's essential frame:
\renewcommand{\labelenumi}{\roman{enumi})}
\begin{enumerate}
\item Local, 
short-time-scale
separation line dynamics are modeled as an Ornstein-Uhlenbeck 
process. (See section 3.)  Here, as depicted in Fig. 2, 
\textit{local} refers to axial
separation line motion in the vicinity of any given
in-nozzle azimuthal angle, $ \phi,$
while the short time scale corresponds to the correlation
time for local axial separation line displacements. 
We will sometimes refer to the latter as the \textit{boundary layer time scale.}
\item Since side load evolution takes place on 
this short time scale, while rocket response to side loading 
occurs on a much longer time scale,
it becomes necessary to derive a coarse grained time correlation
function for local separation line displacement
appropriate to the longer scale. This procedure in turn
leads to 
the crucially important (long-time-scale) side load time correlation function.
See section 4.
\item Given the latter, and focusing
first on rocket rotational dynamics, we show in section 5
that the stochastic
evolution of pitch and yaw likewise correspond to Ornstein-Uhlenbeck
processes. Once this key result is obtained, then theoretical
expressions for altitude-dependent means and variances
of the pitch and yaw rate and pitch and yaw displacement follow. 
Additionally, and as becomes apparent
when interpreting results in section 7,
parametric relationships obtained for associated
(effective) damping and diffusion coefficients 
provide practical guidance 
for minimizing 
the effects of side loads on rotational dynamics.
\item An asymptotic model of translational rocket motion,
appropriate in the practical limit where 
pitch, yaw, and roll all undergo small random variations
about zero, 
follows in section 6.
Again, the simplified linear
model, derived from the general nonlinear model
in \cite{nilabh2010}, allows analytical determination of
altitude-dependent means and variances for the rocket's
lateral velocity and displacement components.
\end{enumerate}

Once a complete model is thus established, 
altitude-dependent variances of pitch and yaw rate, pitch and yaw
displacement, lateral
velocities, and lateral displacements
are computed and compared 
against ensemble average estimates obtained via
the model in \cite{nilabh2010} (section 7).
\section{Physical consistency of Model I}
As detailed in \cite{nilabh2010}, and in response to
the decaying altitude-dependent ambient pressure, the mean position of
boundary layer separation line, $ x_s (t) , $
travels 
down the nozzle axis toward the nozzle exit, with motion taking
place on a relatively slow time scale, $ \tau_a = \Delta x_a /V_R , $
where $ \Delta x_a $ is the characteristic incremental altitude
over which ambient pressure varies and $ V_R $ is the 
characteristic rocket speed.
Superposed on this slow motion
is a fast, random, azimuthally homogeneous stochastic motion. 
Following \cite{nilabh2010},
the joint probability density, $ p_s , $ associated with the instantaneous random 
separation 
line shape
is given by
\begin{equation}\label{pspdf}
p_s( s_1, s_2, \ldots , s_N ) = \prod_I p_I = 
\frac{1}{(2 \pi \sigma_s)^{N/2} } \exp \big[ 
-\frac{ s_1^2 +s_2^2 + s_3^2 + \ldots + s_N^2}{2 \sigma_s^2} \big]
\end{equation}
where, as shown in Fig. 3, $ s_I $ is the random axial
displacement of the separation line at azimuthal angle
$ \phi_I , $ and $ \sigma_s^2 $ is the (assumed) constant
variance of local separation line displacements.

Constituent displacements in the set of $ N $ displacements
are assumed independent, and based on experimental observations
\cite{ref11, ref12, ref17}, gaussian. Thus, each $ p_I $  
is given by
\begin{equation}
p_I (s_I) = \frac{1}{\sqrt{ 2 \pi \sigma_s^2 } } \exp \Big[ - \frac{s_I^2}{2 
\sigma_s^2} \Big]
\end{equation}
In moving to a continuous description of the separation line,
\cite{nilabh2010} assumes that 
\begin{equation}\label{scorrln}
< s (\phi,t) s (\phi',t) > = \sigma_s^2 \delta( \phi -\phi')
\end{equation}

\begin{figure}[!hb]
\centering
\includegraphics[width=4.5in]{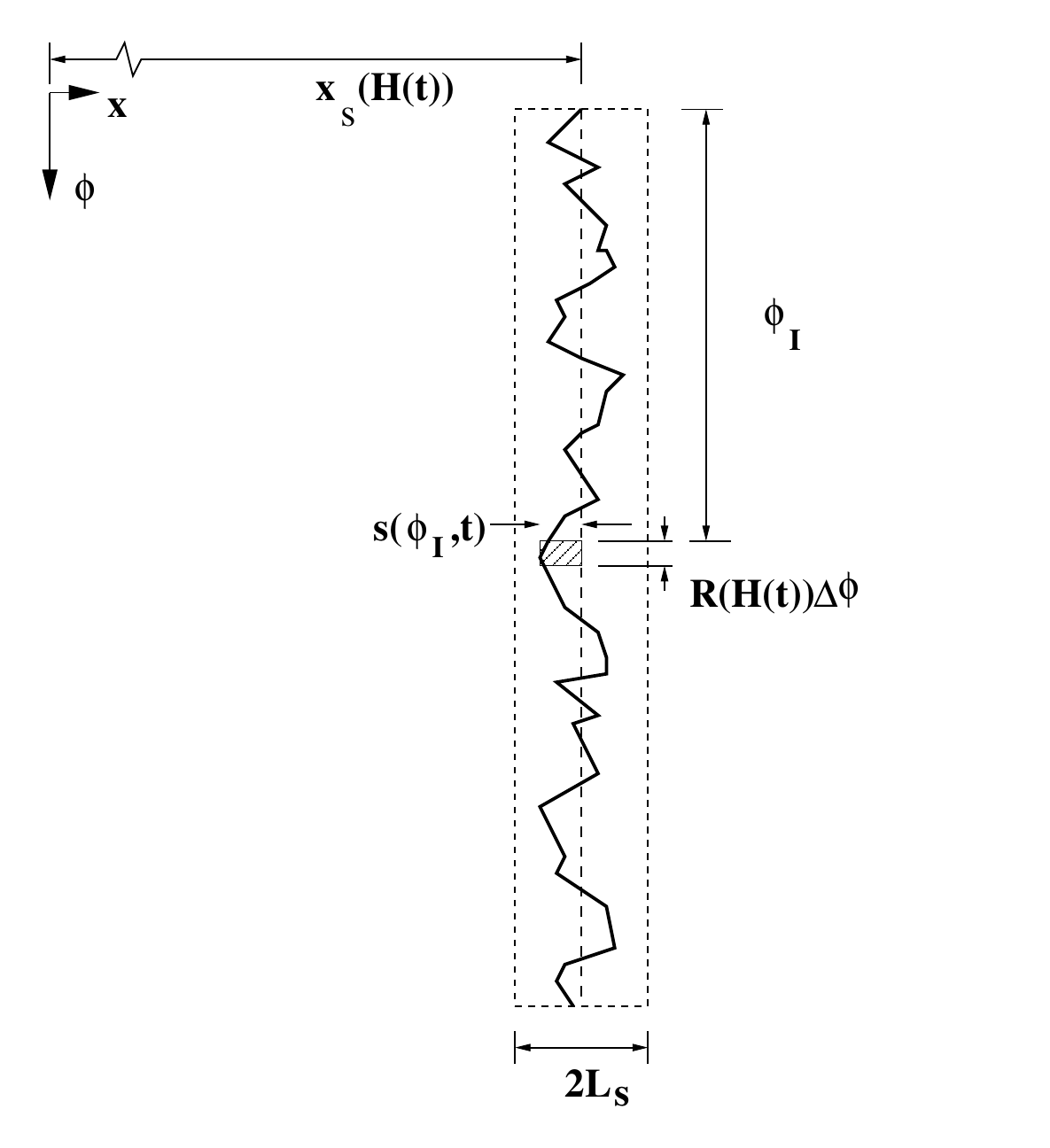}
\caption{Model I \cite{nilabh2010} separation line model.
The mean separation line position, 
$ x_s(H(t)) , $ 
moves down the nozzle axis, 
on the slow time scale associated with 
vertical rocket motion. By contrast, axial separation line motion about
$ x_s(H(t)) , $ 
at any angular position, $ \phi_I , $
is random, and takes place on a much 
shorter time scale. Rapid axial motion, in addition, 
is confined to the nominal
shock-boundary layer interaction zone, again denoted by $ L_s . $
Pressures upstream and downstream
of the instantaneous separation line, $ P_i = P_i(H(t)) $
and $ P_2 = P_2 (H(t)) , $ respectively, are
assumed to be spatially uniform within $ L_s . $
Adapted from \cite{nilabh2010}.}
\end{figure}

Considering next the side load, we express
the instantaneous force vector produced by asymmetric boundary 
layer separation, $ \mathbf{F_s}(t) , $ as a sum of radial and 
axial components
\begin{equation}\label{GrindEQ__26_}
\mathbf{ F_s}(t) = \mathbf{F_r}(t) + \mathbf{F_x}(t)                          
\end{equation}
In \cite{nilabh2010}, the following \textit{ad hoc}
side load model was assumed:
\renewcommand{\labelenumi}{\Alph{enumi})}
\begin{enumerate}
\item  $ F_{sy} $ and $ F_{sz} $ are independent, gaussian random variables,
\item  $< F_{sy} > = 0 $ and $< F_{sz} > = 0 , $
\item  $ \Big< \big(F_{sy}-<F_{sy}>\big)^2 \Big> =\Big< \big(F_{sz}-
<F_{sz}>\big)^2 \Big> = \sigma^2 , $
\end{enumerate}
where, assuming ergodicity, $ < \cdot > $ denotes either an
ensemble or time average, and where the separation line model above
is used to calculate the force variance $ \sigma^2 . $
The side load components $ F_{sy} $ and $ F_{sz} $ are expressed
with respect to rocket-fixed coordinates; see Fig. 4 in section 5 below.

In order to demonstrate the physical consistency
of Model I, we first show that the
\textit{assumed} properties, A) -C), can be \textit{derived} from the simple
separation line model in \cite{nilabh2010}. We then follow \cite{nilabh2010}
and show that the model in A) - C) leads to experimentally
observed side load amplitude and direction densities.

Considering assumption B) first, we calculate:
\begin{equation}\label{fyavg}
\big< F_{sy} (H(t)) \big>_s = R [H(t)] \big[P_i(H(t)) - P_a(H(t)) \big] 
\int_0^{2 \pi}
\sin \phi  \big< s (\phi,t) \big>_s d \phi
\end{equation}
and
\begin{equation}\label{fzavg}
\big< F_{sz} (H(t)) \big>_s = R [H(t)] \big[P_i(H(t)) - P_a(H(t)) \big] 
\int_0^{2 \pi}
\cos \phi  \big< s (\phi,t) \big>_s d \phi
\end{equation}
where $R\left(H(t)\right)$ is the nozzle radius at the axial location of the 
mean separation line at time \textit{t}$,$  
$ {P}_i\left(H\left(t\right)\right)-P_a(H\left(t\right)) = 
\Delta P \left( H(t) \right) , $ is 
the pressure jump across the separation-inducing shock at altitude $H(t) , $
$ P_i $ and $ P_a $ are the wall pressures immediately
upstream and downstream of the instantaneous separation line, and
where we have approximated the downstream pressure as the
instantaneous ambient pressure, $ P_a (H(t)) $ [3,8,15].
See \cite{nilabh2010} for further discussion.

In order to evaluate these averages, express the $ k^{th} $ 
realization of, e.g., $ F_{sz} , $
in discrete form as
\begin{equation}\label{fsum}
F_{sz}^{(k)} ( s_1 , s_2 , \ldots , s_N ) = R [H(t)] \big[P_i(H(t)) - P_a(H(t)) 
\big] \sum_{I=1}^M 
s^{(k)} (\phi_I) \cos \phi_I \Delta \phi
\end{equation}
where $ s^{(k)} (\phi_I) $ is the associated separation line displacement 
at $ \phi_I .$ 
Taking the ensemble average term by term, and noting that
\begin{displaymath}
\big< s_I \big>_s = \int_{- \infty}^{\infty} \int_{- \infty}^{\infty} \ldots 
\int_{- \infty}^{\infty}
s_I p_s (s_1 , s_2 , \ldots , s_N) ds_1 ds_2 \cdots ds_N  = 0 
\end{displaymath}
then confirms B):
\begin{equation}\label{fsumz}
\big< F_{sy} (H) \big>_s = 0 \qquad \qquad  \big< F_{sz} (H) \big>_s = 0
\end{equation}

Turning next to A), since $ F_{sy} $ and $ F_{sz} $ represent sums of say, $ N $ 
independent gaussian random variables, then both are gaussian.
Computing 
\begin{equation*}
<F_{sy}(H) F_{sz}(H) > = R^2(H) \Delta P^2 (H) 
\sigma_s^2 \int_0^{2 \pi} \frac{1}{2}
\sin 2 \phi d \phi = 0
\end{equation*}
where (\ref{scorrln}) has been used,
likewise shows that $ F_{sy} (H) $ and $ F_{sz} (H) $ are independent.

Finally, for C),
since $ <F_{sy}^2(H)> = R^2(H) \Delta P^2 (H) \sigma_s^2 
\int_0^{2 \pi} 
\cos^2 \phi d \phi , $ and \\
$ <F_{sz}^2(H)> = R^2(H) \Delta P^2 (H) \sigma_s^2 
\int_0^{2 \pi} 
\sin^2 \phi d \phi , $ then
\begin{equation}
<F_{sy}^2(H)>= < F_{sz}^2(H)> 
\end{equation}
\subsection{Derivation of density functions
for side load amplitude and direction}\label{secB}
We note two important experimental and numerical observations 
concerning the side load, $ \mathbf{F_s} $ (within rigid, axisymmetric 
nozzles):
\renewcommand{\labelenumi}{\alph{enumi})}
\begin{enumerate}
\item the probability density of the random amplitude, 
$ A =  |\mathbf{F_s}| , $ is a Rayleigh distribution \cite{ref7, ref7a, ref7b}, 
and 
\item the random instantaneous direction, $ \phi_s , $ of $ \mathbf{F_s} $ 
is uniformly distributed over the periphery of the nozzle, or 
$ p_{\phi_s}(\phi_s) = 1 / 2 \pi , $ where $ p_{\phi_s} $ is the pdf of the 
side load direction \cite{ref7, ref7a, ref7b}.
\end{enumerate}

As noted, both observations can be \textit{derived}, starting from 
the simple statistical model of random side 
loads, A) - C), immediately above.
Thus, given $ A $ and $ \phi_s , $ the instantaneous side load components 
in body-fixed \textit{y} and \textit{z} directions (see Fig. 4) are given by 
\begin{displaymath}
F_{sy} = A 
\cos \phi_s  \hspace{2cm}  F_{sz} = A \sin \phi_s
\end{displaymath}

Following \cite{nilabh2010},
write $ F_{sy} $ and $ F_{sz} $ as $ F_{sy}= \bar{Y} = A \cos \phi_s $
and $ F_{sz} = \bar{Z}=A\sin \phi_s ;$ thus, the joint probability density 
associated with $ F_{sy} $ and $ F_{sz} $ can be expressed as
\begin{equation}\label{GrindEQ__27_} 
p_{\bar{Y}\bar{Z}} (\bar{Y},\bar{Z})=p_{\bar{Y}} (\bar{Y}) p_{\bar{Z}} 
(\bar{Z})=\frac{1}{2\pi \sigma^{2} } \exp \left(-\frac{\bar{Y}^{2} 
+\bar{Z}^{2}}{2\sigma^{2} } \right) 
\end{equation} 
Following \cite{nilabh2010}, we restate $p_{\bar{Y}\bar{Z}} $ in terms of $ A $ and  
$ \phi_s $ as, 
\begin{equation}\label{GrindEQ__28_} 
p_{A\phi_s} =\, |J|p_{\bar{Y}\bar{Z}} (\bar{Y},\bar{Z}) 
\end{equation} 
where $ p_{A\phi_s}(A,\phi_s) $ is the joint pdf for the random amplitude and 
direction of $ mathbf{F_r},$ and where the jacobian determinant is given by
\begin{equation}\label{GrindEQ__29_} 
|J|=\left|\begin{array}{cc} {\frac{\partial \bar{Y}}{\partial A} } & 
{\frac{\partial \bar{Y}}{\partial \phi_s } } \\ {\frac{\partial 
\bar{Z}}{\partial A} } & {\frac{\partial \bar{Z}}{\partial \phi_s} } 
\end{array}\right|=A 
\end{equation} 
Thus,
\begin{equation}\label{GrindEQ__30_}
p_{A\phi_s } (A,\phi_s )=\frac{A}{2\pi \sigma^{2} } \exp \left(-\frac{A^{2} 
}{2\sigma^{2} } \right)=\left(\frac{1}{2\pi } 
\right)\left[\frac{A}{\sigma^{2} } \exp \left(-\frac{A^{2} }{2\sigma^{2} } 
\right)\right]=p_{\phi_s } (\phi_s )p_{A} (A) 
\end{equation} 
where,
\begin{equation}\label{GrindEQ__31_} 
p_{\phi_s } (\phi_s )=\frac{1}{2\pi } \hspace{1.5cm} 0<\phi_s \le 2\pi  
\end{equation} 
is the uniform probability density underlying the random direction 
$ \phi_s , $ and 
\begin{equation}\label{GrindEQ__32_} 
p_{A} (A)=\frac{A}{\sigma^{2} } \exp \left(-\frac{A^{2} }{2\sigma^{2} } 
\right) 
\end{equation} 
is the Rayleigh distribution for the amplitude $ A . $ 

In summary, we have shown that the separation line model
in \cite{nilabh2010}: i) allows derivation of the assumed side load
model, A) - C), in \cite{nilabh2010}, and ii) provides a theoretical
basis for explaining experimentally observed \cite{ref7, ref7a, ref7b} 
side load
amplitude and direction densities. 
\section{Ornstein-Uhlenbeck model of separation line 
dynamics}\label{stoch-model-A}
Theoretical determination of rocket response to side loads
requires that the time correlation function for either side load component,
$ \left< F_{s \alpha} (t') F_{s \alpha} (t) \right> , $
be first determined.
This and the next section develops 
$ \left< F_{s \alpha} (t') F_{s \alpha} (t) \right> $
in two steps. First, we propose
(and physically justify) that local 
separation line dynamics can be modeled as an Ornstein-Uhlenbeck process.  
Once this assumption is made, then the second step
rests on a rigorous argument 
showing that on the relatively long rocket dynamics time scale, the boundary 
layer separation line shape, and importantly, associated side load 
components, are all delta correlated in time. See section 4.   

We propose the following simple, explicit stochastic 
model of separation line dynamics:
\begin{equation} \label{GrindEQ__48_} 
ds_i(t)=-ks_i(t)+\ \sqrt{D_s}dW(t) 
\end{equation} 
where $s_i \left(t\right)=s\left({\phi }_i,t\right)$ is the instantaneous 
separation line position at ${\phi }_i$, \textit{k} and $D_s$ are damping and 
effective diffusion coefficients, and $dW(t)$ is a differential Weiner 
process.  This equation, describing an Ornstein-Uhlenbeck process, allows 
straightforward, physically consistent calculation of statistical properties 
associated with separation line motion and, more importantly, serves as the 
first link in a chain that connects short-time-scale random 
separation line motion to short-time-scale random side loads, and in turn, to 
long-time-scale stochastic rotational rocket dynamics.  

The form of this equation is chosen based on the following experimental 
features, observed in shock-separated flows near compression corners and 
blunt fins:
\renewcommand{\labelenumi}{\alph{enumi})}
\begin{enumerate}
\item Under statistically stationary conditions, the feet of 
separation-inducing shocks oscillate randomly, up- and downstream, over 
limited distances, about a fixed mean position; see, e.g., \cite{ref11,
ref12, ref17}.
\item As observed in \cite{ref12,ref17}
the distribution of shock foot positions within 
the shock-boundary layer interaction zone is approximately gaussian.
\item The time correlation of shock foot positions, as indicated by 
wall pressure measurements within the shock-boundary layer interaction zone, 
decays rapidly for time intervals, $\triangle t$, larger than a short 
correlation time, ${\tau }_s,\ $a feature that can be inferred, for example, 
from \cite{plotkin}.
\end{enumerate}
Physically, the damping term captures the fact that the shock sits within a 
pressure-potential energy well. Thus, downstream shock excursions 
incrementally decrease and increase, respectively, upstream 
and downstream shock 
face pressures; the resulting pressure imbalance forces the shock back 
upstream. A similar mechanism operates during upstream excursions.  
Introduction of a Weiner process models the combined random forcing produced 
by advection of turbulent boundary layer structures through the upstream side 
of the shock foot and pressure oscillations emanating from the downstream 
separated boundary layer and recirculation zone.  

We note that the proposed 
model is qualitatively consistent with 
Plotkin's model of boundary layer-driven shock motion near compression 
corners and blunt fins \cite{plotkin}.  Plotkin's model, which captures low 
frequency spectra of wall pressure fluctuations within these flows, 
corresponds to a generalized Ornstein-Uhlenbeck process in which a 
deterministic linear damping term is superposed with a non-Markovian random 
forcing term.  We use an ordinary OU process model, incorporating a Wiener 
process, since again, it is consistent with the above observations and more 
particularly, since it allows much simpler calculation of statistical 
properties.  
\section{Derivation of the coarse grained side load
time correlation function}
Given the model in Eq. (\ref{GrindEQ__48_}), our path shifts 
to obtaining a mathematically consistent description of resulting side load 
statistics.  To accomplish this, we take advantage of the significant 
separation in time scales that exists between large-scale, low frequency 
random separation line motion \cite{ref11, ref12, ref17}, and the 
relatively slow dynamics of the 
rocket.  Thus, define the long time scale as $\tilde{t}=t\ \epsilon^{-1},$  
where $\epsilon << 1$ is a dimensionless scale factor and \textit{t} is the 
(short) time scale associated with low frequency, large scale separation line 
motion.  The long time scale can be chosen to correspond to any of a number 
of dynamical features; the chosen scale determines $ \epsilon $. Since 
we are interested in the dynamical response of the rocket, 
we choose $\tilde{t}$ to be on the order of $ \tau_R 
={L V_R^{-1}},$where $ L $ and $ V_R $ are the length and characteristic 
speed of the rocket. Thus, 
$ \epsilon=f^{-1}{\tau_R^{-1}} , $  where \textit{f} is a characteristic 
frequency from the low frequency band associated with large scale shock foot 
motion \cite{ref11, ref12}.  Based on the rocket
parameters given in \cite{nilabh2010},
${\tau }_R$ is on the order 
of $2({10}^{-2})$ s; thus, 
since $f^{-1}$ is on the order of ${10}^{-3}$ s 
\cite{ref12, ref17}, $ \epsilon=O\left({10}^{-2}\right)$ to 
$O\left({10}^{-1}\right)$.  

Using well known expressions \cite{gardiner}
for the mean and variance of an OU process, 
one can readily show that the 
process becomes stationary on time scales that are 
long relative to the characteristic period for large scale separation line 
motion, i.e., for $t\gg k^{-1} .$ 
Under stationary conditions, the variance becomes 
independent of time, $var\left(s_i\left(t\right)\right)=D_s\ 
/\left(2k\right),\ $ and the time correlation can be placed in the form:
\begin{equation} \label{GrindEQ__49_} 
<s_i\left(\tau \right)s_i\left(0\right)>_s=\ \frac{D_s{\tau }_c}{2}\ {\rm 
exp} (-\left|\tau \right|\ /\ {\tau }_c\ ) 
\end{equation} 
where $\tau =t_2-t_1$, and ${\tau }_c=\int^{\infty 
}_0{<s_i}\left(t\right)s_i\left(0\right)>dt/\ 
var\left(s_i\left(t\right)\right),$ is the correlation time.  
Under stationary conditions, ${\tau }_c=k^{-1}$ and 
$var\left(s_i\left(t\right)\right)=D_s{\tau }_c\ /\ \ 2\ ={\sigma }^2_s$.   

Given the above time correlation, we can now show that on 
long time scales,$\ \tilde{t},$  $<s_i \left(\tau 
\right)s_i\left(0\right)>_s,$ approaches delta-function like character.  
First, define a normalized time correlation function as 
\begin{equation} \label{GrindEQ__50_} 
R_s \left(\tau \right)=\frac{<s_i\left(\tau 
\right)s_i\left(0\right)>_s}{2{\sigma }^2_s{\tau }_c} 
\end{equation} 
where $\int^{\infty }_{-\infty }{R_s }\left(\tau \right)d\tau $=1. Next, 
define a rescaled version of $ R_s \left(\tau \right)$ as ${R_{s \epsilon} }
\left(\widetilde{\tau }\right)={\varepsilon }^{-1} R_s \left({\varepsilon 
}^{-1}\tau \right),$ or
\begin{equation} \label{GrindEQ__51_} 
R_{s \varepsilon} \left(\widetilde{\tau 
}\right)=\frac{<s_i\left(\widetilde{\tau 
}\right)s_i\left(0\right)>_s}{2{\sigma }^2_s{\tau }_c\varepsilon }\  
\end{equation} 
where $\widetilde{\tau }={\epsilon}^{-1}\tau $. Finally, for small $\varepsilon $, 
$ R_{s \varepsilon }\left(\widetilde{\tau }\right)\to \delta 
\left(\widetilde{\tau }\right).\ $ Thus, since $\varepsilon \ll 1$, 
the long-time-scale
behavior of the time correlation function can be stated as
\begin{equation} \label{GrindEQ__52_} 
<s_i\left(\widetilde{\tau }\right)s_i\left(0\right)>_s=2{\sigma }^2_s{\tau 
}_c\varepsilon \ \delta \left(\widetilde{\tau }\right) 
\end{equation} 
Thus, from this point on, we
focus on time scales that are on the order of the rocket-dynamics 
time scale, ${\tau }_R$, drop the tilde on $\tilde{t} , $ and specify      
\begin{equation} \label{GrindEQ__53_} 
<s \left( \varphi',\ t' \right) s \left( \varphi ,t \right)>_s = 2 \sigma^2_s  
\tau_c \varepsilon \ \delta \left( \varphi'-\varphi \right) \delta 
\left(t'-t \right) 
\end{equation} 
as the long-time-scale space-time correlation for $s(\varphi ,t),$  
where again we assume delta correlated statistics in 
the angular direction \cite{nilabh2010}.

Given this correlation, the long-time-scale time correlations for the 
side load components can be finally calculated by
combining and averaging instantaneous, 
non-averaged versions of Eqs. \eqref{fyavg} and \eqref{fzavg}:
\begin{equation} \label{GrindEQ__54_} 
\ <F_{sy} ( 
t^{'} )F_{sy}(t)>_s=<F_{sz} (t^{'} )F_{sz}\left(t\right)>_s=R(t{'})\ 
R (t) \ \triangle P(t^{'})\ \triangle P(t)\ 2\pi 
\sigma^2_s \tau_c \varepsilon \delta (t'-t) 
\end{equation} 

Obtaining the long-time-scale side load time correlation
function represents a crucial step since it allows calculation of
side load effects on rocket rotational and translational dynamics.
\section{Pitch and yaw rate response to 
random side loading}\label{stoch-model-
B}
Given 
$ < F_{sy} (t') F_{sy}(t) > = < F_{sz}(t') F_{sz}(t) >  , $ derived above,
we now show that during the period when side loads act, $\ 0<t\le T$, 
the long-time-scale evolution of the rocket's pitch and yaw rates correspond 
to Ornstein-Uhlenbeck processes.  
Given this key result, altitude-dependent means and variances
for the rocket's pitch and yaw rates, both during the side load period
and after, can be determined.
\subsection{Derivation of Ornstein-Uhlenbeck rotational dynamics}
As a necessary point of reference, we first
note the equations governing
rotational dynamics \cite{nilabh2010}:
\begin{equation} \label{GrindEQ__22_} 
I_{xx} \dot{\omega }_{x} +\left(\dot{I}_{xx} +\frac{2}{5} |\dot{M}|R_{e}^{2} 
\right)\omega _{x} =0 
\end{equation} 
\begin{equation} \label{GrindEQ__23_} 
I\dot{\omega }_{y} +(I_{xx} -I)\omega _{x} \omega _{z} +\dot{I}\omega _{y} 
+|\dot{M}|[(L-b)^{2} +0.25R_{e}^{2} ]\omega _{y} -\frac{|\dot{M}|LR_{R}^{4} 
}{10v_{e} R_{e}^{2} } \omega _{x} \omega _{z} =\mathbf{M}_{ext} .\mathbf{j} 
\end{equation} 
\begin{equation} \label{GrindEQ__24_} 
I\dot{\omega }_{z} -(I_{xx} -I)\omega _{x} \omega _{y} +\dot{I}\omega _{z} 
+|\dot{M}|[(L-b)^{2} +0.25R_{e}^{2} ]\omega _{z} +\frac{|\dot{M}|LR_{R}^{4} 
}{10v_{e} R_{e}^{2} } \omega _{x} \omega _{y} =\mathbf{M}_{ext} .{\kern 1pt} 
\, \mathbf{k} 
\end{equation} 
Regarding notation, moments of inertia are evaluated with respect to rocket-
fixed coordinates, $ I = I(t) $ and $ I_{xx} = I_{xx}(t) $
are, respectively,
the moments of inertia with respect to 
either lateral coordinate, $ z $ or $ y , $ and
the longitudinal rocket axis, $ R_R , $ $ L , $ $ b , $ and $ R_e , $
are, respectively, rocket radius, rocket length
and half-length, and nozzle exit radius,
$ v_e $ is the speed of the exiting flow from the nozzle,
$ \dot{M} $ is the associated mass flux, and 
$ \omega_x , $ $ \omega_y , $ and 
$ \omega_z  $ are, respectively, 
the roll, yaw, and pitch rates.   Unit vectors 
$ \mathbf{i}, \mathbf{j}, \mathbf{k}, $ are those of the rocket-fixed system.
See \cite{nilabh2010} for a full description and Fig. 4
for a schematic of the rotational and translational variables
describing rocket motion.

\begin{figure}[!hb]
\centering
\includegraphics[width=5.5in]{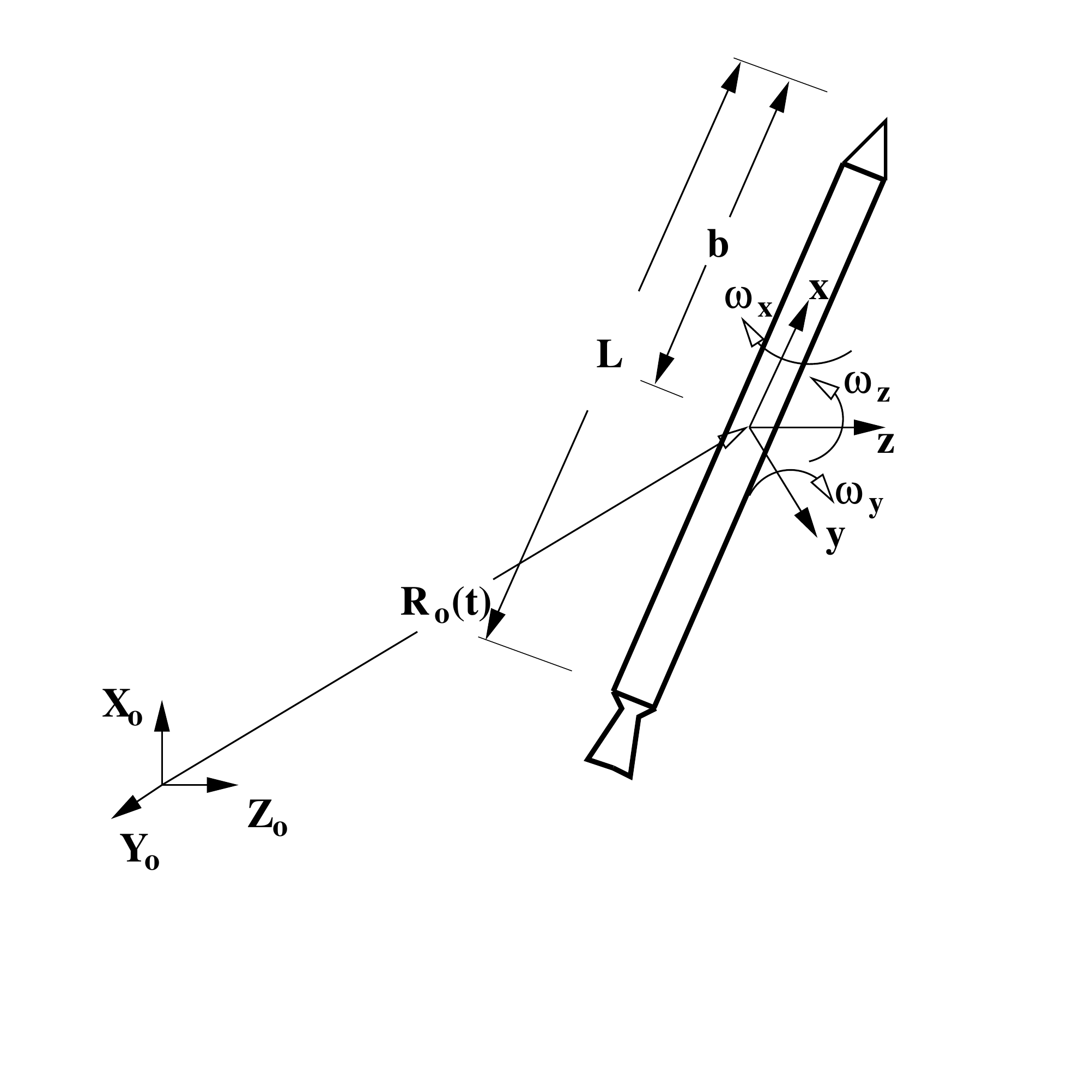}
\caption{Rocket model. Roll, pitch, and yaw rates are shown, respectively,
as $ \omega_x , $ $ \omega_z , $ and $ \omega_y ; $
center of mass position is given by
$ R_o (t)= [ X_o (t), Y_o (t) , Z_o (t) ] , $ 
while rocket-fixed coordinates are $ [x,y,z] . $
See \cite{nilabh2010} for further details.}
\end{figure}

Since the line of action of the aerodynamic load is assumed coincident with the 
rocket's longitudinal axis, 
the only force that contributes to 
the external moment, $\mathbf{M}_{ext} , $ in Eqs. 
\eqref{GrindEQ__23_} and 
\eqref{GrindEQ__24_} is the in-nozzle stochastic side load. The moment due to 
these stochastic side loads can thus be computed as
\begin{equation} \label{GrindEQ__25_} 
\mathbf{M}_{ext} =-(L-b+x_s(t))\mathbf{i}\, {\rm x}\, (F_{sy} \mathbf{j}+F_{sz} 
\mathbf{k}) 
\end{equation} 

As an important aside, and
as a means of isolating side load effects, the present model,
as well as that in \cite{nilabh2010}, 
does not incorporate wind loading.
A simple scaling argument (see Appendix
B) indicates that random winds: i) under most conditions,
do not excite rotational motion, and ii) simply function as 
an \textit{additive}
source of variance in the rocket's \textit{translational} motion.  
In other words, wind appears to have minimal
influence on the
stochastic, altitude-dependent evolution of \textit{rotational} 
dynamics. Rather, 
(launch-site-specific) mean and random winds simply
produce 
whole-rocket, random, \textit{lateral translational}
motion, superposed on a 
deterministic translational drift. 

During the short side load period, and as found by scaling, it is observed from 
the equation governing roll, Eq. \eqref{GrindEQ__22_}, that the coefficients 
on ${\dot{\omega}}_x$ and ${\omega }_x$, $I_{xx}$ and 
$(\dot{I}+{{\frac{2}{5}}}\left|\dot{M}\right|R^2_e)$, respectively, remain 
essentially constant.  
[As shown in \cite{nilabh2010},
for the model considered, the side load period, \textit{T}, 
is approximately 10.85 s.]
Thus, for an assumed initial roll rate, ${\omega 
}_x\left(0\right)=0,$  
and during the side load period, $0\le t\le T,$ 
the rate of roll remains zero, ${\omega }_x$= 0.
Indeed, as shown in \cite{nilabh2010}, and due to the absence of
roll-inducing torques, this feature holds throughout 
any given simulated flight. 

Thus, we focus on the evolution of yaw and pitch,
as determined by 
Eqs. \eqref{GrindEQ__23_} and \eqref{GrindEQ__24_}, respectively, 
and observe
that these can be placed in the forms:
\begin{equation} \label{GrindEQ__55_} 
d{\omega }_y=-A(t){\omega }_y(t)dt+B(t)F_{sz}\left(t\right)dt 
\end{equation} 
\begin{equation} \label{GrindEQ__56_} 
d{\omega }_z=-A(t){\omega }_z(t)dt-B(t)F_{sy}\left(t\right)dt 
\end{equation} 
where 
\begin{equation} \label{GrindEQ__57_} 
A(t)=[\dot{I}+|\dot{M}|[\left(L-b)^2+0.25R^2_e\right]/\ I 
\end{equation} 
is a positive damping coefficient (positive as determined by a 
straightforward order of magnitude analysis) and 
\begin{displaymath}
B(t)=\left(L-b\right)/\ I(t)
\end{displaymath}
Next, from, e.g., \cite{gardiner}, we note that gaussian, zero mean random 
functions that are delta correlated in time, such as $F_{sz}(t)$ and 
$F_{sy}\left(t\right),$ can be related to Weiner processes via 
\begin{equation} \label{GrindEQ__58_} 
B\left(t\right)F_{sz}\left(t\right)dt=\sqrt{D(t)}\ 
dW_z\left(t\right)=\sqrt{D(t)}{[W}_z\left(t+dt\right)-W_z\left(t\right)] 
\end{equation} 
where $W_z\left(t\right)$ is a Wiener process associated with 
$F_{sz}\left(t\right)$, \textit{D(t)} is a diffusion coefficient, and where a 
similar expression holds for $F_{sy}\left(t\right)$.   Here, using 
\begin{equation} \label{GrindEQ__59_} 
\iint^{t+\triangle 
t}_t{B(t")B(t')}<F_{sz}\left(t"\right)F_{sz}\left(t^{'} \right)>_sdt"dt'=\iint^{
t+\triangle t}_t \sqrt{D(t")} \sqrt{D(t')} {dW_z(t")dW_z(t')}=D(t) \triangle t 
\end{equation} 
it is readily shown that 
\begin{equation} \label{GrindEQ__60_} 
D\left(t\right)={(L-b)^2I}^{-2}\left(t\right)R^2{\left(t\right){\triangle 
}^2P(t)\ 2\pi {\sigma }^2_s{\tau }_c\varepsilon } 
\end{equation} 
where, consistent with dimensional requirements, $D(t)$ has units of $t^{-
3}$. 

Thus, we arrive at the important result that
during the period $0\le t\le T$, the random rotational dynamics of the 
rocket correspond to Ornstein-Uhlenbeck processes: 
\begin{equation}\label{GrindEQ__61_}
d{\omega }_y=-A(t){\omega }_y(t)dt+\sqrt{D(t)}dW_z\left(t\right)              
\end{equation}
\begin{equation} \label{GrindEQ__62_} 
d{\omega }_z=-A(t){\omega }_z(t)dt-\sqrt{D(t)}dW_y\left(t\right) 
\end{equation} 
We discuss in section 7 the physical features underlying the
competition between damping, as embodied by the terms involving $ A(t)$
in \eqref{GrindEQ__61_} and \eqref{GrindEQ__62_},
and amplification of pitch and yaw rates, 
as captured by the $ D(t) $ terms.
\subsection{Pitch and yaw rate response}
While in many problems, the damping and diffusion coefficients are constant, 
here they are not.  Nevertheless, Eqs. \eqref{GrindEQ__61_} and 
\eqref{GrindEQ__62_} can be integrated (for individual realizations) via an 
integrating factor: 

\begin{equation} \label{GrindEQ__63_} 
{\omega }_{\alpha }\left(t\right)={\rm exp}{\rm (}-
\int^t_0{A\left(s^{'}\right)ds^{'})\int^t_0{\left\{\sqrt{D\left(s^{'}\right)}{\rm 
exp}?{\rm (}\int^{s^{'}}_0{A(s"{\rm )ds"\ )}}\right\}\ dW\left(s^{'}\right)}{\rm 
\ }} 
\end{equation} 
where the subscript $\alpha $ denotes either $y$ or $z$. Given Eq. 
\eqref{GrindEQ__63_}, the mean yaw and pitch rates can be determined:

\[{<\omega }_{\alpha }\left(t\right)>={\exp  \left(-
\int^t_0{A\left(s^{'}\right)ds^{'}}\right)\ 
}\int^t_0{\left\{\sqrt{D\left(s^{'}\right)}{\exp}{\rm 
(}\int^{s^{'}}_0{A(s"{\rm )ds"\ )}}\right\}<dW\left(s^{'}\right)}>\] 
or since $<dW\left(s^{'}\right)>=0,$ 

\begin{equation} \label{GrindEQ__64_} 
{<\omega }_{\alpha }\left(t\right)>=0  \hspace{1.5cm}              0\le t<T 
\end{equation} 
[Note, we are abusing standard notation by expressing the expectation over 
Wiener processes, $E_{\omega,s'} W\left(s'+ds'\right)-W(s')$ as 
$<dW (s^{'})>.$] Likewise, \eqref{GrindEQ__63_} allows determination 
of the time-dependent variances for pitch and yaw rates:
\begin{eqnarray}
<\omega^2_{\alpha }(t)>= & \exp  \left(-
2\int^t_0 A(s')ds'\right) \ 
\int^t_0 \int^t_0 \left\{ \sqrt{D(s')D(p')}{\rm exp} \left( 
\int^{s'}_0 A(s" )ds" \ \right) {\rm exp} \left( \int^{p'}_0 A(p" )dp" \ 
\right) \right\} <d{\beta }_{s'p'}> \nonumber \\
 = & \int^t_0{\int^t_0{\left\{F(s',p')\right\}}<dW\left(s'\right)dW\left(p'
\right)>} \nonumber
\end{eqnarray}
where $<d{\beta }_{s'p'}>=<dW\left(s'\right)dW\left(p'\right)>.$  
Expressing the term 
$\int^t_0{\int^t_0{\left\{F(s',p')\right\}}<dW\left(s'\right)dW\left(p'
\right)>}$ 
as a discrete double sum and using  $\triangle W_i\triangle 
W_j=\triangle t{\delta }_{ij},$ it is readily shown that 
\begin{equation} \label{GrindEQ__65_} 
<{\omega }^2_{\alpha }\left(t\right)>={\exp  \left(-
2\int^t_0{A\left(s'\right)ds'}\right)\ }\int^t_0{\left\{{\rm D} ({{\rm 
s}}^{'}) {\rm \ exp} {\rm (2}\int^{s'}_0{A(s"{\rm )ds"\ 
)}}\right\}ds'}   \hspace{1cm}   0\le t<T 
\end{equation} 
It is also easily shown that in the case of constant $A$ and $D$, this 
formula leads to the well-known \cite{gardiner} variance expression for 
constant coefficient OU processes. 

During the post-side load period, $T\le t\le T_f,$ the 
(short-time-scale) side load terms in 
equations \eqref{GrindEQ__55_} and \eqref{GrindEQ__56_}
no longer appear, and thus
neither do associated (long-time-scale) Wiener processes 
in Eqs. \eqref{GrindEQ__61_} and 
\eqref{GrindEQ__62_}.
Thus, carrying out a single  
realization integration of the latter two equations, we obtain
\begin{equation} \label{GrindEQ__66_} 
{\omega }_{\alpha }\left(t\right)={\omega }_{\alpha }\left(T\right)
{\exp  
\left[ - \int_T^t A(q) dq \right]\ } 
\end{equation} 
where ${\omega }_{\alpha }\left(T\right)\ $is a random initial condition on 
the post-side load pitch and yaw rate evolution equations.  Since 
$<{\omega }_{\alpha }\left(T\right)>$ = 0, it is clear that 
\begin{equation} \label{GrindEQ__67_} 
<{\omega }_{\alpha }\left(t\right)> = 0\ \ \ \ \ \ \ \ \ \ \ \ \ \ T<t\le T_f 
\end{equation} 
consistent with numerical results in \cite{nilabh2010}.

Likewise, forming $<{\omega }^2_{\alpha }\left(t\right)>$ and using 
\begin{equation}\label{GrindEQ__68_}
<{\omega }^2_{\alpha }\left(T\right)>={\exp  \left(-
2\int^T_0{A\left(s'\right)ds'}\right)\ }\int^T_0{\left\{{\rm D}{{\rm 
s}}^{{\rm '}}{\rm \ exp}{\rm (2}\int^{s'}_0{A(s"{\rm )ds"\ 
)}}\right\}ds'}
\end{equation} 
obtained from Eq. \eqref{GrindEQ__65_}, the time-dependent variance of 
yaw and pitch rates during the post-side-load period follow as:
\begin{equation} \label{GrindEQ__69_} 
<{\omega }^2_{\alpha }\left(t\right)>={{<{\omega }^2_{\alpha }\left(T\right)> 
{\exp \left[ -2 \int_T^t A(q) dq \right]\ }}}   \hspace{.3cm} T<t\le T_f 
\end{equation} 

In closing this subsection, we again
note that the rocket's rate of roll,$\ {\omega 
}_x , $ is uncoupled from moments produced by random side loads (see Eq. (23)); 
this is reflected in the solution of Eq. \eqref{GrindEQ__22_}:
\begin{equation} \label{GrindEQ__70_} 
{\omega }_x\left(t\right)={\omega }_x\left(0\right){\rm exp} 
[\int^t_0{F\left(t'\right)dt']} 
\end{equation} 
where $F\left(t'\right)=[{\dot{I}}_{xx}+{{\frac{2}{5}}}|\dot{M}$\textbar 
$R^2_e]{\dot{I}}^{-1}_{xx}.$  Thus, for an assumed initial roll rate of zero, 
roll rate remains zero, ${\omega }_x\left(t\right)=0$, throughout any given 
flight.  Again, computed results in \cite{nilabh2010} 
are consistent with this 
observation. 
\subsection{Pitch and yaw angle response}\label{stoch-model-C}
Here, time-dependent
pitch and yaw angle variances are determined, as well as the time correlation
function, $ \left< \psi_{\alpha} (t') \psi_{\alpha} (t) \right> , $
where $ \psi_{\alpha} $ represents either the pitch or yaw angle.
Determining the evolution of pitch/yaw variances provides
a further consistency check 
between theory and numerical experiments,
and, in addition, provides important physical insight into
the large, almost explosive growth in translational
velocity and displacement variances described in section 
\ref{stoch-model-D} below.
The time correlation function is needed in order to determine
these lateral translational responses.

During the side load period, single realization
evolution of pitch and yaw angles follows by
integration of \eqref{GrindEQ__63_}:
\begin{equation}\label{model-D-eqn-1}
\psi_{\alpha}(t)=\int_0^t \exp f(q) G_{\alpha} (q) dq
\end{equation}
where 
\begin{equation}\label{model-D-eqn-2}
f(q) = - \int_0^q A(q") dq"
\end{equation}
and 
\begin{equation}\label{model-D-eqn-3}
G_{\alpha} (q) = 
- \int_0^q \left[ \sqrt{D(s')} \exp \int_0^{s'} A(s") ds" \right] dW_{\alpha} (s')
\end{equation}

Taking the expectation over the Weiner process, $ d W_{\alpha} (s') , $
we find, consistent with numerical experiments,
that mean pitch and yaw angles remain fixed at zero throughout the side load
period:
\begin{equation}\label{model-D-eqn-4}
\left<\psi_{\alpha}(t) \right> = 0    \hspace{1.3cm} 0 \le t < T
\end{equation}

Evolution of pitch/yaw angle variance
during $ 0 \le t < T $ is given by
\begin{equation}\label{model-D-eqn-5}
\left<\psi_{\alpha}^2(t) \right> =\int_0^t \int_0^t \exp f(q') \exp f(q) 
\left< G_{\alpha} (q') G_{\alpha} (q) \right> dq' dq
\end{equation}
or, using 
\begin{displaymath}
\left< \int_0^{t_1} F(s') dW(s') \cdot \int_0^{t_2} F(s") dW(s") \right> =
\int_0^{min(t_1,t_2)} F^2(s') ds',
\end{displaymath}
\begin{equation}\label{model-D-eqn-6}
\left<\psi_{\alpha}^2(t) \right> =\int_0^t \int_0^t \exp f(q') \exp f(q) 
\left( \int_0^{min(q,q')} D(s') \left[ exp 2 \int_0^{s'} 2 A(s") ds" \right] 
ds' \right) dq' dq
\end{equation}
which, by symmetry, becomes:
\begin{equation}\label{model-D-eqn-7}
\left<\psi_{\alpha}^2(t) \right> = 2 \int_0^t \left[ \int_0^{q'} \exp f(q') \exp f(q) 
\left( \int_0^{q} D(s') \left[ \mathrm{exp} \ 2 \int_0^{s'} 2 A(s") ds" \right] 
ds' \right) dq \right] dq'
\end{equation}
The detailed form of \eqref{model-D-eqn-7} appropriate to the
present rocket model is given in Appendix A.

Following the side load period, pitch/yaw 
evolution follows by integration of \eqref{GrindEQ__66_}:
\begin{equation}\label{model-D-eqn-8}
\psi_{\alpha}(t)=\omega_{\alpha}(T) \int_T^t h(s') ds' 
+  \psi_{\alpha}(T) 
\end{equation}
where
\begin{equation}\label{model-D-eqn-8A}
h(s') = \exp \left( -\int_T^{s'} A(s") ds"  \right)
\end{equation}
Taking the expectation again shows that
average pitch/yaw angles remain at zero throughout the post-side-load
period:
\begin{equation}\label{model-D-eqn-9}
\left<\psi_{\alpha}(t) \right> = 0    \hspace{1.3cm} T \le t \le T_f
\end{equation}
where, by 
\eqref{GrindEQ__63_},
$ \left< \omega_{\alpha}(T) \right> = 0 .$ Again, this is 
consistent with results of numerical experiments in \cite{nilabh2010}.

Post-side-load variance follows directly from \eqref{model-D-eqn-8}:
\begin{eqnarray}\label{model-D-eqn-10}
\left< \psi_{\alpha}^2(t) \right> =& \left< \omega_{\alpha}^2(T) \right> 
\left( \int_T^t h(s') ds' \right)^2 +
2 \left< \omega_{\alpha}(T) \psi_{\alpha}(T) \right> 
\left( \int_T^t h(s') ds'  \right)  + \nonumber \\ 
 & + \left< \psi_{\alpha}^2 (T) \right>
  \hspace{1.3cm} T \le t \le T_f
\end{eqnarray}
where the correlation
$ \left< \omega_{\alpha}(T) \psi_{\alpha}(T) \right> $ is non-zero and is
given by
\begin{equation}\label{model-D-eqn-11}
\left< \omega_{\alpha}(T) \psi_{\alpha}(T) \right> =
\int_0^T \exp f(q) \exp f(T) 
\left< G_{\alpha} (q) G_{\alpha} (T) \right> dq
\end{equation}
or
\begin{equation}\label{model-D-eqn-12}
\left< \omega_{\alpha}(T) \psi_{\alpha}(T) \right> =
\int_0^T \int_0^q \int_0^T \exp f(q) \exp f(T) \sqrt{D(s')} \sqrt{D(s")}
g(s') g(s") \left< dW_{\alpha}(s') dW_{\alpha} (s") \right> dq
\end{equation}
or finally by
\begin{equation}\label{model-D-eqn-13}
\left< \omega_{\alpha}(T) \psi_{\alpha}(T) \right> =
\int_0^T \exp f(q) \exp f(T) \int_0^q D(s')
g^2(s') ds' dq 
\hspace{.3cm} T \le t \le T_f
\end{equation}
Here, $ g(s') = \exp \left( \int_0^{s'} A(p) dp \right) ,$
and $ f $ and $ G_{\alpha} $ are given, respectively, by 
\eqref{model-D-eqn-2} and 
\eqref{model-D-eqn-3}.  The detailed  problem-specific 
form of \eqref{model-D-eqn-13} is again given in Appendix A. 
\section{Asymptotic rocket response to small-
amplitude stochastic pitch and yaw}\label{stoch-model-D}
Under conditions where the rocket experiences zero-mean 
stochastic side loading, we anticipate that the Euler angles $ \phi(t),\ \ 
\theta (t),$ and $\psi (t)\ $all undergo small, random variations about zero.  
Under these conditions, a leading order asymptotic model of rocket lateral 
velocity and displacement response can be used to validate and interpret 
the numerically estimated statistics obtained via Model I.

Thus, considering the full translational equations of motion given in
\cite{nilabh2010}, we observe that under conditions where 
\begin{equation} \label{GrindEQ__71_} 
\left[\phi \left(t\right),\theta \left(t\right),\psi 
\left(t\right)\right]=O(\epsilon_o ) 
\end{equation} 
for all $t$, where $\epsilon_o \ll 1 ,$ these equations assume the forms:
\begin{equation}\label{GrindEQ__73_} 
M \ddot{X}_{o} =(P_{e} -P_{a} )A_{e} +|\dot{M}|v_{e} -0.5C_{D} A_{R} 
\rho_{a} (\dot{X}_{o}^{2} +\dot{Y}_{o}^{2} +\dot{Z}_{o}^{2} )-Mg+O(\epsilon_o )           
\end{equation}
\begin{equation}\label{GrindEQ__73A_} 
M \ddot{Y}_{o} = F_T \psi + F_{sy} +2|\dot{M}|(L-b)\omega _{z} +O(\epsilon_o ) 
\end{equation} 
\begin{equation} \label{GrindEQ__74_} 
M \ddot{Z}_{o} =F_T \theta + F_{sz} -
2|\dot{M}|(L-b)\omega _{y} +O(\epsilon_o ) 
\end{equation} 
where
\begin{displaymath}
F_T = (P_e -P_a) A_e + |\dot{M}| v_{ex} 
\end{displaymath}
is the total thrust force,
and where
$\phi (t), $ $ \theta(t) , $ and $ \psi(t) $ correspond, respectively,
to roll, pitch, and yaw angles. Other new terms include
the rocket mass, $ M = M(t) , $ the nozzle exit pressure and exit area, $ P_e $
and $ A_e , $ the (mach number dependent) drag coefficient, $ C_D , $ the 
rocket cross-sectional area, $ A_R , $ the altitude-dependent
ambient density, $ \rho_e , $ and the rocket center-of-mass
position $ [X_o(t), Y_o(t), Z_o(t) ] . $
See Fig. 4, and 
refer to \cite{nilabh2010} for
a detailed description. 

An important note: scaling, as well as results below show
that the thrust terms (involving $ F_T ) $
in the lateral equations of motion, \eqref{GrindEQ__73A_} and
\eqref{GrindEQ__74_}, must be included;
due to the magnitude
of $ F_T , $ even small, order $ \epsilon $
pitch and yaws can produce non-negligible, and indeed, dominant lateral
forces.
In addition, due to the restriction of $\phi (t)$ to small magnitudes, the 
following development does not hold when the rocket is given a non-zero roll.  
Note too another abuse of notation in the use of dimensional terms in the 
order symbols.

Due to the importance of minimizing random lateral 
velocities and displacements, and due to
the miniscule effects of side loads on vertical motion
\cite{nilabh2010},
we focus on rocket dynamics in the lateral 
(i.e., $y-$ and $z-$) directions.   The above equations 
show that during the side load period, $0\le t<T,$ leading order
lateral translational dynamics are determined
by the summed effects of three random forces: 
\renewcommand{\labelenumi}{\alph{enumi})}
\begin{enumerate}
\item the nozzle side load,
$ F_{s \eta} , $ 
\item  the 
lateral thrust component, $ \pm F_T \psi_{\alpha \pm} , $
produced by random variations in pitch and yaw
angles, and
\item a second reaction term, 
$ \pm 2|\dot{M}|(L-b)\omega _{\alpha \pm} , $
produced by incremental (random) changes in the direction
of the mass flux vector, $ \mathbf{\dot{M} } . $
\end{enumerate}
[Here, $ \psi_{\alpha +} = \psi , $ the yaw angle, 
$ \psi_{\alpha -} = \theta , $ the pitch angle, 
$ \omega_{\alpha +}={\omega}_z, $ the pitch rate, and ${\omega }_{\alpha -}={\omega }_y, $ 
the yaw rate.]
Following the side load period, 
only the latter two random forcing terms continue to act.

Letting $\eta (t)$ represent either $Y_o(t)$ or $Z_o(t)$, and focusing in 
turn on the side load and post-side-load periods, one can integrate equations 
\eqref{GrindEQ__73A_} and \eqref{GrindEQ__74_} to obtain single realization 
solutions.  Thus, over the side load period, 
\begin{equation} \label{GrindEQ__75_} 
\dot{\eta }\left(t\right)= 
\int^t_0{\frac{F_{s\eta }(s')}{M(s')}ds'}\ \pm 
2\left(L-
b\right)\int^t_0{\frac{\left|\dot{M}\left(s'\right)\right|}{M\left(s'\right
)}}{\omega }_{\alpha \pm }\left(s'\right)ds' 
\pm \int_0^t \frac{F_T(s') \psi_{\alpha \pm}(s')}{M(s')}  ds' 
\quad   0\le t<T 
\end{equation} 
while over the post-side-load period
\begin{equation} \label{GrindEQ__76_} 
\dot{\eta }\left(t\right)=\dot{\eta }\left(T\right) 
\pm 2\left(L-
b\right)\int^t_T{\frac{\left|\dot{M}\left(s'\right)\right|}{M\left(s'\right
)}}{\omega }_{\alpha \pm }\left(s'\right)ds'  
\pm \int_T^t \frac{F_T(s') 
\psi_{\alpha \pm}}{M(s')} ds' 
\quad   \ T\le t\le T_f 
\end{equation} 
[Use Eqs. \eqref{GrindEQ__73A_} and \eqref{GrindEQ__74_} to choose the signs in 
\eqref{GrindEQ__75_} and \eqref{GrindEQ__76_}.]

In order to determine single realization displacements, first define  
\begin{equation} \label{GrindEQ__77_} 
G_{\eta }\left(t\right)=\int^t_0{\frac{F_{s\eta }(s')}{M(s')}ds'} 
\end{equation} 
\begin{equation} \label{GrindEQ__78_} 
H_{\eta }\left(t;t_o\right)=2\left(L-
b\right)\int^t_{t_o}{\frac{\left|\dot{M}\left(s'\right)\right|}{M\left(s'
\right)}}{\omega }_{\alpha \pm }\left(s'\right)ds' 
\end{equation} 
\begin{equation} \label{GrindEQ__78A_} 
\tilde{F} (t;t_o) = \int_{t_o}^t \frac{F_T(s') 
\psi_{\alpha \pm}(s')}{M(s')} ds'
\end{equation}
and integrate \eqref{GrindEQ__75_} and \eqref{GrindEQ__76_} to obtain:
\begin{equation} \label{GrindEQ__79_} 
\eta \left(t\right)=\int^t_0{G_{\eta }\left(s'\right)ds'}\pm 
\int^t_0{H_{\eta }\left(s';0\right)ds'} \pm \int_0^t \tilde{F} (s';0) ds'
  \quad       0\le t<T 
\end{equation} 
\begin{equation} \label{GrindEQ__80_} 
\eta \left(t\right)=\dot{\eta }\left(T\right)(t-T)\pm \int^t_T{H_{\eta 
}\left(s';T\right)ds'}  \pm \int_T^t \tilde{F} (s';T) ds'
  \quad    T\le t<T_f 
\end{equation} 

Given Eqs. \eqref{GrindEQ__75_} and \eqref{GrindEQ__76_}, time-dependent 
ensemble average lateral velocities are readily calculated:
\[<\dot{\eta }\left(t\right)>=\int^t_0{\frac{<F_{s\eta 
}\left(s'\right)>}{M(s')}ds'}\ \pm 2\left(L-
b\right)\int^t_0{\frac{\left|\dot{M}\left(s'\right)\right|}{M\left(s' \right
)}}{<\omega }_{\alpha \pm }\left(s'\right)>ds' \pm 
\int_0^t \frac{ F_T(s') <\psi_{\alpha \pm} (s')>}{M(s')} ds' \quad    0\le t<T\] 
\[<\dot{\eta }\left( t\right)>=<\dot{\eta }\left(T\right)>\pm 2\left(L-
b\right)\int^t_T{\frac{\left|\dot{M}\left(s'\right)\right|}{M\left(s'\right
)}}{<\omega }_{\alpha \pm }\left(s'\right)>ds'  \pm
\int_T^t \frac{F_T(s') <\psi_{\alpha \pm} (s')>}{M(s')} ds' \quad   T\le t\le T_f\] 
From Eqs. \eqref{fsumz}, \eqref{GrindEQ__64_} and 
\eqref{GrindEQ__67_}, it is clear that ensemble average lateral velocity 
components remain zero throughout the entire flight period:
\begin{equation} \label{GrindEQ__81_} 
<\dot{\eta }\left(t\right)>\ =0   \hspace{1.3cm}           0\le t\le T_f 
\end{equation}   
a result that is again consistent with numerical experiments \cite{nilabh2010}.

Variances of the lateral velocity components likewise follow from Eqs. 
\eqref{GrindEQ__75_} and \eqref{GrindEQ__76_}:
\begin{eqnarray}\label{GrindEQ__82_}
<\dot{\eta}^2 \left(t\right)>= & \int^t_0 \int^t_0 \left< \frac{F_{s\eta} (s')}{M(s')}
\frac{F_{s \eta }(p')}{M(p')} \right> ds'dp' +4 \left(L-
b\right)^2 \int^t_0 \int^t_0 \frac{\left|\dot{M}\left(s'\right)\right|}{M
\left(s'\right)} \frac{\left|\dot{M}\left(p'\right)\right|}{M\left(p' \ 
\right)}<{\omega }_{\alpha \pm }\left(s'\right){\omega }_{\alpha \pm 
}\left(p'\right)>ds'dp' + \nonumber \\
 & + \int^t_0 \int^t_0 \frac{{F}_T(s') {F}_T(p')}{M(s')M(p')} <\psi_{\alpha \pm}(s')
\psi_{\alpha \pm} (p')> ds' dp' + \nonumber \\
 & + 2 (L-b) \int_0^t \int_0^t \frac{|\dot{M} (s')|}{M(s')} \frac{F_T(p')}{M(p')}
\left< \psi_{\alpha \pm} (p') \omega_{\alpha \pm} (s') \right> ds' dp'
\end{eqnarray}
for 0$\le t<T$, and where it is assumed that pitch and yaw rates are 
uncorrelated with side loads,
\noindent  $<{\omega }_{\alpha \pm }\left(s'\right)F_{s\alpha 
}\left(p'\right)>\ =0. $ 

Similarly, over the post-side-load period, $T\le 
t\le T_f$,
\begin{eqnarray}\label{GrindEQ__83_} 
<{\dot{\eta }}^2\left(t\right)>= & <{\dot{\eta }}^2\left(T\right)>+
4{\left(L-b\right)}^2\int^t_T{\int^t_T{\frac{\left|\dot{M}\left(s'\right)\right|}
{M\left(s\right)}}\frac{\left|\dot{M}\left(p'\right)\right|}
{M\left(p' \ \right)}}<{\omega }_{\alpha \pm }\left(s'\right)
{\omega }_{\alpha \pm }\left(p'\right)>ds'dp'  + \nonumber \\
 & +  \int^t_T \int^t_T \frac{{F}_T(s') {F}_T(p')}{M(s')M(p')} <\psi_{\alpha \pm}(s')
\psi_{\alpha \pm} (p')> ds' dp' + \nonumber \\
 & + 2 (L-b) \int_T^t \int_T^t \frac{|\dot{M} (s')|}{M(s')} \frac{F_T(p')}{M(p')}
\left< \psi_{\alpha \pm} (p') \omega_{\alpha \pm} (s') \right> ds' dp'
\end{eqnarray} 
where we assume that $<\dot{\eta }\left(T\right)\int^t_T {\frac{\left|\dot{M}\left(s'\right)\right|}
{M\left(s' \right)}}{\omega }_{\alpha \pm }\left(s'\right)ds'>\ =\ 0$.
See Appendix A for problem-specific versions of \eqref{GrindEQ__82_}
and \eqref{GrindEQ__83_}.

Ensemble averages and variances for lateral displacements can be obtained 
using the single realization solutions in Eqs. \eqref{GrindEQ__79_} and 
\eqref{GrindEQ__80_}.  Thus, it is again readily shown that average lateral 
displacements are zero both during and after the side-loading period:
\begin{equation} \label{GrindEQ__88_} 
<\eta \left(t\right)>=\int^t_0{<G_{\eta }\left(s'\right)>ds'}\pm 
\int^t_0{<H_{\eta }\left(s';0\right)>ds'} \pm \int_0^t \tilde{F}(s') ds' =0   
\hspace{.8cm}     0\le t<T 
\end{equation} 
\begin{equation} \label{GrindEQ__89_} 
<\eta \left(t\right)>=<\dot{\eta }\left(T\right)>(t-T)\pm \int^t_T{<H_{\eta 
}\left(s';T\right)>ds'} \pm \int_T^t \tilde{F}(s')ds' =0  \hspace{.8cm}    T\le t<T_f 
\end{equation} 
which is again consistent with \cite{nilabh2010}.
Note, $G_{\eta } ,$ $H_{\eta }, $ and $ \tilde{F} $ are given by Eqs. \eqref{GrindEQ__77_}, 
\eqref{GrindEQ__78_} and \eqref{GrindEQ__78A_}, respectively.  

Variances are likewise found:      
\begin{eqnarray}\label{GrindEQ__90_} 
<{\eta }^2\left(t\right)>=& \int^t_0{\int^t_0{<G_{\eta }
\left(s'\right)G_{\eta }(p')>}ds'dp'+\int^t_0{\int^t_0{<H_{\eta }\left(s' ; 0\right)
H_{\eta }(p';0)>}ds'dp'}}  + \nonumber \\
 & + \int_0^t \int_0^t <\tilde{F} (s';0) \tilde{F}(p';0)> ds' dp'  \hspace{1.0cm} 0\le t<T 
\end{eqnarray} 
\begin{eqnarray}\label{GrindEQ__91_} 
<{\eta }^2\left(t\right)>= &<{\dot{\eta }}^2\left(T\right)>{\left(t-
T\right)}^2+\int^t_T{\int^t_T{<H_{\eta }\left(s';T\right)H_{\eta 
}\left(p';T\right)>}ds'dp'} +  \nonumber \\
 & + \int_T^t \int_T^t <\tilde{F} (s';T) \tilde{F}(p';T)> ds' dp'  
 \hspace{1cm} T\le t<T_f 
\end{eqnarray} 
where it is assumed that all cross-correlations between $ G_{\eta} ,$
$ H_{\eta} , $ $ \tilde{F} , $ and $ \eta(T) $ are zero.
Problem-specific versions of 
\eqref{GrindEQ__90_} and
\eqref{GrindEQ__91_} are given in Appendix A.
\section{Results and discussion}\label{results}
As an important preliminary, we list essential explanatory remarks
and observations. 
\renewcommand{\labelenumi}{\Alph{enumi})}
\begin{enumerate}
\item As shown in Appendix A, 
most working variance formulae involve one or more time-dependent
terms: $ \Delta P(t) , $ $ R(t) , $ $ M(t) , $ and $ I(t). $
Here, these (non-stochastic) terms are determined using 
numerical data from execution of Model I \cite{nilabh2010}.
In cases where the current model (Model II) is used to predict
the dynamics of actual rockets, $ R(t) , $
the time-dependent nozzle radius corresponding to the
instantaneous mean separation line position, $ x_s(t) , $
represents the most difficult-to-determine parameter. At least two
straightforward approaches, i.e., methods not requiring 
high-level modeling and computation, are available.
The first \cite{keanini}, combines
one of several semi-empirical separation
pressure correlations, see, e.g., \cite{ref1}, with
a model of nozzle flow upstream of separation.
See Fig. 9 in \cite{keanini}.
The second uses scaling
to obtain an approximate model of mean separation
line motion as a function of time (altitude).
This approach will be described in a future paper.

\item The parameter $ \Delta P (t) = P_i(t) -P_a(t) , $
requires estimation of $ P_i (t) , $ the nozzle-wall pressure
near the incipient separation point.  Again, an alternative to
complex modeling and numerics can be found in a semi-empirical
approach outlined in \cite{keanini}.

\item Examination of the working formulae
used to compute theoretical variances (see Appendix A)
shows that the three \textit{a priori} unknown boundary layer parameters,
$ \sigma_s , $ $ \epsilon , $ and $ \tau_c ,$
representing, respectively, the nominal length of the boundary
layer-shock interaction zone $(\sigma_s \approx L_s ) , $
the ratio of the boundary layer to 
rocket dynamics time scales, and the correlation time
for local separation line displacements,
\textit{only} appear as the product, 
$ \kappa_o= \sigma^2 \tau_c \epsilon . $
Thus, Model II has only \textit{one} available adjustable 
parameter, $ \kappa_o , $
for fitting theoretical to experimental data.

\item Rather than employing any of a number of
standard fitting procedures for estimating
$ \kappa_o , $ we use the following simple approach.
First, express $ \kappa_o $ as $ \kappa_o = \beta_o \sigma_o^2
\epsilon_o \tau_{co} , $  where $ \beta_o $ becomes the fitting
parameter and the last three parameters
are assigned 
nominal, empirically- or scaling-based values: 
$ \sigma_o = 2.54 \ (10^{-2}) \ \mathrm{m} $ \cite{ref1,
ref11},
$ \epsilon = 0.1 $ (scaling), and $ \tau_c = 10^{-3} \ \mathrm{s} $  
\cite{ref11, ref12}. Second,
use straightforward trial and error to estimate $ \beta_o . $
Since the $ \beta_o $ thus obtained is nearly equal to
1/4, we arbitrarily set $ \beta_o = 0.25 . $ 

\item It is important to
note that if we reasonably (though arbitrarily) 
take $ \tau_c $ as the most
uncertain of the three parameters $ \sigma , $ $ \epsilon , $
and $ \tau_c , $ i.e., express $ \tau_c $ as $ \tau_c =  \beta_o \tau_{co} , $
and interpret $ \tau_c $ as the approximate
frequency of large-scale 
separation line motion \cite{ref17},
then the \textit{estimated} value of $ \tau_c^{-1} = 
\mathrm{400} \ \mathrm{Hz}, $
is comparable to 
frequencies $ (\approx \ \mathrm{300 \  Hz} ) $  
experimentally observed in shock-separated flat plate boundary layers
\cite{ref11}.

\item As noted in Appendix A, scaling, as well as 
numerical results (Model I),
show that during the side load period, $ 0 \le t < T, $
variance growth in lateral 
rocket displacements, $ Y_o (t)  $ and $ Z_o (t) , $
remains negligible relative to that observed during
the post-side-load period, $ T < t \le T_f . $
For simplicity, during $ 0 \le t < T , $
theoretical (Model II) displacement variances 
are thus set to $ 0 . $ 

\item Parameters defining the model rocket can be found in \cite{nilabh2010}.

\item An ensemble of 100 numerically simulated rocket 
ascents are used to estimate experimental (Model I) means and variances.
Since time-dependent statistics estimated using a smaller set of 40 ascents
differed by no nore than 14 \% from these, no attempt has been made to examine
larger ensembles.

\item As noted in section 6, theoretical 
time evolution of the following
\textit{mean} values: 
pitch and yaw rate, 
$ <\omega_z (t) > $ and $ <\omega_y (t)> , $ 
pitch and yaw displacement, $ < \theta (t) > $ and $ < \psi(t) > , $ 
lateral velocity, $ < \dot{Y}_{o} > $ and $ < \dot{Z}_{o} > , $ and 
lateral displacement, $ < Y_o(t) > $ and $ < Z_o(t) > , $ 
all remain identically zero
throughout the simulated flight period,
$ 0 \le t \le T_f . $
These results are, in every instance, consistent
with the numerical observations in 
\cite{nilabh2010}.  
\end{enumerate}

In discussing the results, we follow three threads:
\renewcommand{\labelenumi}{\Alph{enumi})}
\begin{enumerate}
\item Evolution of pitch and yaw \textit{rate} variances
turn out to play a central role in evolution of not only pitch
and yaw \textit{displacement} variances, but also
in variance evolution of lateral
translational velocities and displacements.
In the latter cases, \textit{thrust components} produced by
random pitch and yaw emerge as the dominant
mechanism generating translational stochasticity.
\item We find that physical interpretation of many of the results
below can be usefully framed in terms of the 
OU damping and diffusion coefficients, $ A(t) $ and $ D(t) , $
given respectively by \eqref{GrindEQ__57_} and \eqref{GrindEQ__60_}.
The utility of these parameters traces to 
the central role played by random
pitch and yaw rates in stochastic rocket dynamics.
\item The transparent physical content of $ A(t) $ and $ D(t) $
allows straightforward identification of practical approaches for
minimizing or mitigating against the 
effects random pitch and yaw.
\end{enumerate}
\subsection{Pitch and yaw rate response}
Numerical and theoretical pitch and yaw rate variances,
$ <\omega_z^2 (t) > $ and $ <\omega_y^2 (t)> , $ 
are compared in Fig. 5.  
General theoretical expressions,
applicable over the side load period and post load period, 
are given respectively
by Eqs. 
\eqref{GrindEQ__65_} 
and
\eqref{GrindEQ__69_}; corresponding expressions specific to the present model
are given by 
\eqref{app_eqn_1} 
and
\eqref{app_eqn_3}.
It is clear that theoretical pitch and yaw rate variances
remain qualitatively consistent with 
numerical estimates throughout the entire flight period.

\begin{figure}[!hb]
\centering
\includegraphics[width=4.75in]{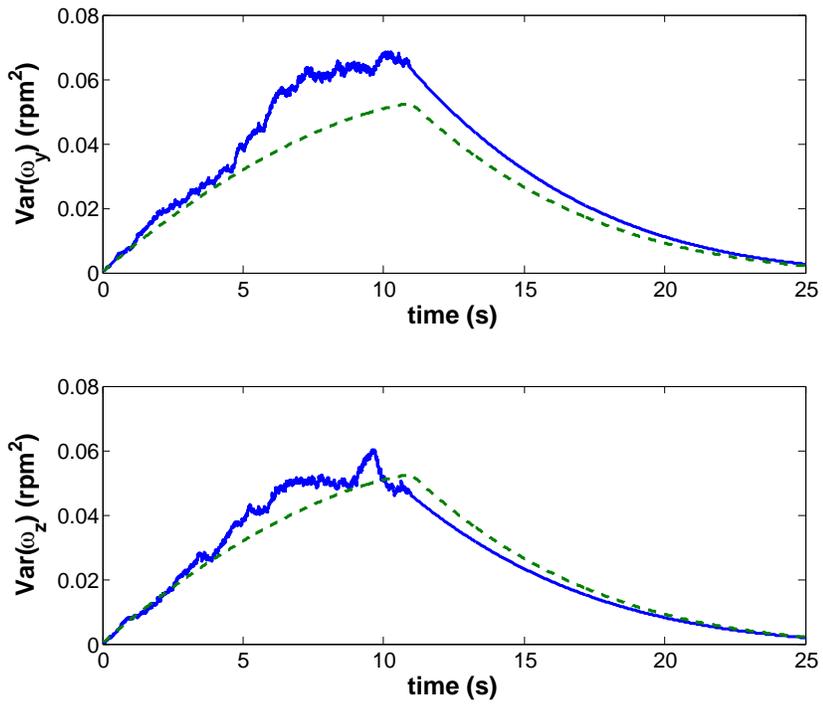}
\caption{Comparison of yaw, $ \omega_y , $ and 
pitch, $ \omega_z , $ rate variance
evolution as estimated using Model I (solid line)
and as computed via Model II (dashed line). The side load period
ends at $ t=10.85 \ \mathrm{s} .$ }
\end{figure}
\subsubsection{Pitch and yaw rate response: side load period}
During the side load period, theoretical and numerically observed
stochasticity, indicated by the variances,
clearly increase in a qualitatively consistent manner.
[Note, 
differences between ensemble-averaged
pitch and yaw rate variances decrease when the ensemble is enlarged
from 40 to 100 simulated flights. We anticipate that the observed
differences in figure 5 become negligible, as they should by symmetry,
as $ N \rightarrow \infty . $ A similar comment applies to all
other numerical results.] 

Physical insight into variance growth during the side load period, and
identification of
design approaches for
minimizing growth of stochastic
pitch and yaw,
follows by looking at a simpler model. Thus, take
the damping and diffusion coefficients, $ A(t) $ and $ D(t) , $
in \eqref{GrindEQ__57_} and \eqref{GrindEQ__60_},
as constant.  Although an approximation here,
this simplification becomes increasingly accurate as the side load period
shortens. Under this approximation,
variance growth follows
the standard OU formula \cite{gardiner}:
\begin{equation}\label{ouguy}
< \omega^2_{\alpha} (t)> = \frac{D_o}{2 A_o} \left( 1 - \exp 
\left( -2 A_o t \right) \right) 
\end{equation}
with the associated rate of variance growth, valid in the limit
$ A_o T << 1, $ given by
\begin{equation}\label{vargrowth}
\frac{d < \omega^2_{\alpha } (t)>}{dt} = D_o + O(A_o T) 
\end{equation} 
Here, $ A_o = A(0) $ and $ D_o = D(0) . $

We use \eqref{ouguy} to examine physical features
underlying pitch/yaw rate
variance in the limit, $ A_o T >> 1 , $ appropriate 
under conditions where damping is strong and/or
ascent is slow.
Equation \eqref{vargrowth} is used to examine
the opposite limit, $ A_o T << 1 , $ i.e., the limit
applicable to the present numerical \cite{nilabh2010} and theoretical models.

Considering first $ A_o T >> 1 , $
the asymptotic variance
takes the form:
\begin{equation}\label{asymp-var}
< \omega_{\alpha}^2 (t) > \approx  \frac{D_o}{2 A_o}  \approx \frac{R^2(0) \Delta^2 P(0) 
\pi \sigma^2 \epsilon}{I(0) |\dot{M}| } 
\end{equation} 
where, as determined by scaling, 
the relationships, 
$ | \dot{M}| (L-b)^2 >> \dot{I} $
and $ | \dot{M}| (L-b)^2 >> 0.25 R_e^2 ,$
have been used; refer to \eqref{GrindEQ__57_}. 

Physically, the right side of \eqref{asymp-var} represents
the ratio of stochasticity \textit{amplification} via side-load-induced
random torques, to stochasticity \textit{stabilization}
via mass flux damping. The former feature becomes apparent
when we restore the squared moment arm, $ (L-b)^2 , $ cancelled
in obtaining \eqref{asymp-var}.

The role of mass flux damping can be ascertained by  
sketching the incremental (vector) change,
$ \boldsymbol{\Delta} \mathbf{\dot{M} } , $
in the mass flux vector, 
$ \mathbf{\dot{M} } , $
produced by an
incremental pitching
or yawing displacement (about the pitch or yaw axis).
Such a sketch shows that
$ \boldsymbol{\Delta} \mathbf{ \dot{M} } $ acts (nominally)
in the plane of the nozzle exit, in a direction opposing the angular
motion.  Thus, 
with reference to
\eqref{GrindEQ__57_} , \eqref{GrindEQ__61_}, and \eqref{GrindEQ__62_},
we recognize that 
$ \boldsymbol{\Delta} \mathbf{ \dot{M} } $
produces a retarding, i.e., stabilizing torque that 
opposes \textit{all} pitch and yaw motions.

Based on this interpretation, 
we can use
\eqref{asymp-var} to identify practical approaches, appropriate when 
$ A_o T >> 1 , $
for either reducing random pitch/yaw amplification,
or increasing mass flux damping:
\renewcommand{\labelenumi}{\alph{enumi})}
\begin{enumerate}
\item reduce the nozzle size (characterized by $ R(0) ),$
\item reduce the near-exit, ground-level pressure difference,
$ \Delta P (0) , $ e.g., by reducing the degree of
over-expansion,
\item increase the rocket's moment of inertia, as characterized by
$ I(0) , $ and/or
\item increase the mass flux, $ \dot{m} . $
\end{enumerate}
These observations presume, reasonably,
that the
boundary layer separation parameters, 
$ \sigma_s , $ $ \tau_c , $ and $ \epsilon $
remain, in an order of magnitude sense,
independent of nozzle size and in-nozzle flow conditions.

Turning to the limit where the side load
period is too short for the asymptotic regime
to set in, $ A_o T << 1 , $ again the limit appropriate
to the present study and that in \cite{nilabh2010},
we expose features that determine
the \textit{rate} of variance growth.
Again using initial values 
for the parameters in \eqref{GrindEQ__60_}, Eq. \eqref{vargrowth}
yields (to $O(A_o T) ) $
\begin{eqnarray}\label{doeqn}
\frac{ d < \omega^2_{\alpha } (t)>}{dt} & = & D_o  \nonumber \\ 
 & = & (L-b)^2 I^{-2}(0) R^2 (0) \Delta^2 P (0) 2 \pi \sigma_s^2 
\tau_c \epsilon  \nonumber \\
\end{eqnarray} 

Again
assuming that the parameters 
$ \sigma_s , $ $ \tau_c , $ and $ \epsilon $
remain nominally independent of nozzle size and in-flight
nozzle flow conditions,
we observe that the rate of pitch/yaw rate variance growth 
can be minimized, e.g., by: 
\renewcommand{\labelenumi}{\alph{enumi})}
\begin{enumerate}
\item again using high-rotational inertia rocket designs, 
\item moving the center of mass toward the
nozzle exit (thus reducing the moment arm, $ L - b  ),$ 
while maintaining 
high $ I , $  
\item using smaller (radius) nozzles, and of course when feasible, 
\item operating in an underexpanded condition, 
\end{enumerate}
where d) eliminates separation and thus side loading.

Other approaches, designed
to reduce $ \sigma_s , $ $ \tau_c , $ and/or $ \epsilon , $
might include boundary layer manipulation via,
e.g., wall fluid injection or suction, and/or active
or passive mechanical forcing. 
\subsubsection{Pitch and yaw rate response: post-side-load period} 
Side loads cease when the rocket reaches an altitude
where ambient pressure drops below
the near-exit nozzle pressure.
From this instant onward, 
and as is apparent in
both numerical and theoretical plots of 
$ < \omega_{\alpha }^2 > $
in Fig. 5,
rocket pitch and yaw,
become subject only to the damping torque produced by
incremental changes in $ \mathbf{\dot{M}} . $

Equation \eqref{GrindEQ__69_} shows that the post-side-load
decay in pitch/yaw rate variance is exponential, with the rate of damping
increasing with time.  Physically, and as shown
by \eqref{GrindEQ__57_}, 
accelerated rotational damping, in turn, reflects the
inverse dependence of the damping coefficient, $ A(t) , $
on the time-decaying moment of inertia, $ I(t) : $
as the rocket becomes decreasingly resistant
to pitch/yaw rotations, the damping torque
can effect ever-larger influence on pitch and yaw.

We can use \eqref{GrindEQ__57_} to quickly identify
design strategies that, e.g., enhance post-side-load
damping of random pitch and yaw. Thus, assuming that
the term in $ R_e^2 $ is small, the case here, then 
\renewcommand{\labelenumi}{\alph{enumi})}
\begin{enumerate}
\item increasing $ \dot{M} , $ and/or 
\item increasing the moment arm, $ L-b ,$ \\
will improve damping.
\end{enumerate}
Clearly, trade-offs are required
since, for example, strategy b) \textit{enhances}
variance growth during the side load period.
By contrast, increasing nozzle mass flux, at least
up to magnitudes for which $ |\dot{I}| $ remains 
smaller than $ |\dot{M}| (L-b)^2 , $
is always
beneficial since this enhances mass-flux damping.
[Interestingly, for large enough $ |\dot{M}| , $
the  negative term, $ \dot{I} , $ in \eqref{GrindEQ__57_}
can become dominant, transforming $ A(t) $ 
into an \textit{amplification} coefficient. A simple
analog that explains this effect
can be found in the gravity-driven pendulum:
shortening the pendulum length during motion
increases the amplitude of the motion.]
\subsection{Pitch/yaw displacement variance}
Pitch and yaw displacement variances are compared in 
Fig. 6. 
General theoretical expressions for the
side load and post side load periods
are given respectively
by Eqs.
\eqref{model-D-eqn-7} 
and
\eqref{model-D-eqn-10}, and problem-specific versions are
derived in Appendix A.
As in the case of pitch/yaw rate variances,
reasonable consistency
between theoretical and numerical results is again observed.

\begin{figure}[!hb]
\centering
\includegraphics[width=4.75in]{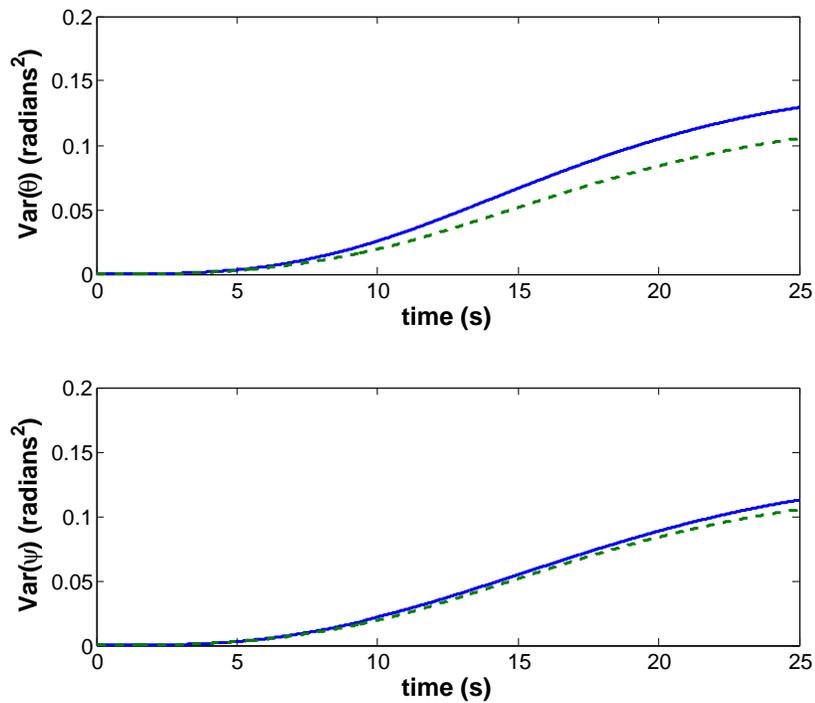}
\caption{Comparison of pitch, $ \theta , $ and yaw,
$ \psi , $ displacement variance
evolution as estimated using Model I (solid line)
and as computed via Model II (dashed line). The side load period
ends at $ t=10.85 \ \mathrm{s} .$ }
\end{figure}
 
In order to interpret variance growth during the side load
period, we again note that due to the brevity
of the side load period, 
$ A(t) $ and $ D(t) $ 
can be approximated
as constant and equal
to, say, $ A_o = A(0) $ and $ D_o = D(0) . $ Starting with
\eqref{model-D-eqn-7}, 
noting that $ A_o T << 1 , $
and expanding 
\eqref{model-D-eqn-7} in $ A_o t $
then leads to
\begin{displaymath}
< \psi_{\alpha}^2 (t) > \approx \frac{D_o}{3} t^3 + O(A_o^4 t^4) 
\end{displaymath}
where 
$  1 \ >> \ \frac{D_o}{3} t^3 >> A_o^4 t^4 , $
and where $ D_o $ is given by \eqref{doeqn}.
Thus, the observed growth in pitch/yaw variance 
during side loading reflects the dominant effect of
\textit{diffusive}, i.e., stochastic, side-load-driven growth
over mass-flux damping.
Variance growth is slow since the diffusion coefficient, $ D_o , $
is small; a quick order of magnitude estimate,
yielding $ D_o \approx 10^{-5} \ \mathrm{s^{-3}} , $
shows that $ D_o t^3 /3 $ is on the order
of observed and predicted side load period 
variances.

During the post-side-load period, and based on 
\eqref{model-D-eqn-8}, 
it proves useful to interpret the term,
$ \int_T^t h(s') ds' , $
appearing in 
\eqref{model-D-eqn-10} as a
response function; here, the function
yields the total random change in pitch/yaw displacement, 
over the interval from $ T $ to $ t , $
produced by the random initial
pitch/yaw rate, $ \omega_{\alpha} (T) . $ 
Thus, examining the three terms on the right side of
\eqref{model-D-eqn-10}, we observe that the displacement
variance at $ t , $ 
$ < \psi_{\alpha}^2 (t)> , $ corresponds
to the sum of: i) the average squared response
to the random input
$ \omega_{\alpha} (T) , $ ii) the \textit{weighted} average \textit{linear}
response to $ \omega_{\alpha} (T) , $ where weighting is with respect
to the random initial displacement, $ \psi_{\alpha}(T) , $ and
iii) the initial displacement variance, 
$ < \psi_{\alpha}^2 (T)> . $
While the response function increases with time,
$ t - T , $ it is readily shown that the rate of increase
decays with increasing
$ t - T ; $ 
thus, in Fig. 6, we observe
a roll-off in displacement variance.

It now becomes apparent that post-side-load displacement
variance at any time $ t $ increases (decreases) 
with any increase (decrease) in any
of the above inputs. This picture provides an explanation, for example,
of the larger post-side-load \textit{numerical} (Model I) yaw variances
observed in Fig. 6.  As another example,
under circumstances where damping,  $ A(t) , $ decreases,
we expect, based on simple physical intuition, 
that the lag between angular velocities and displacements
decreases and thus, the correlation 
$ < \omega_{\alpha} (T) \psi_{\alpha} (T) > , $ increases.
Thus, consistent with our intuition,
weakly damped rockets
arrive at $ t=T $ with a greater range of random
pitch/yaw displacements; this initial input, in turn, leads, via
the response function, to larger post-side-load displacement
variance. Clearly, practical design considerations, similar
to those discussed above,  can be identified and
used to manipulate the response function.
\subsection{Lateral velocity and displacement variance}
Lateral velocity and displacement variances obtained via Models I and II are 
compared in Figs. 7 and 8, respectively.
Full theoretical velocity expressions 
applicable to the
side load and post side load periods
are given respectively
by Eqs. 
\eqref{GrindEQ__82_}
and
\eqref{GrindEQ__83_}. 
Corresponding expressions for theoretical displacement variances
are given in 
\eqref{GrindEQ__90_}
and
\eqref{GrindEQ__91_}. 

\begin{figure}[!hb]
\centering
\includegraphics[width=4.75in]{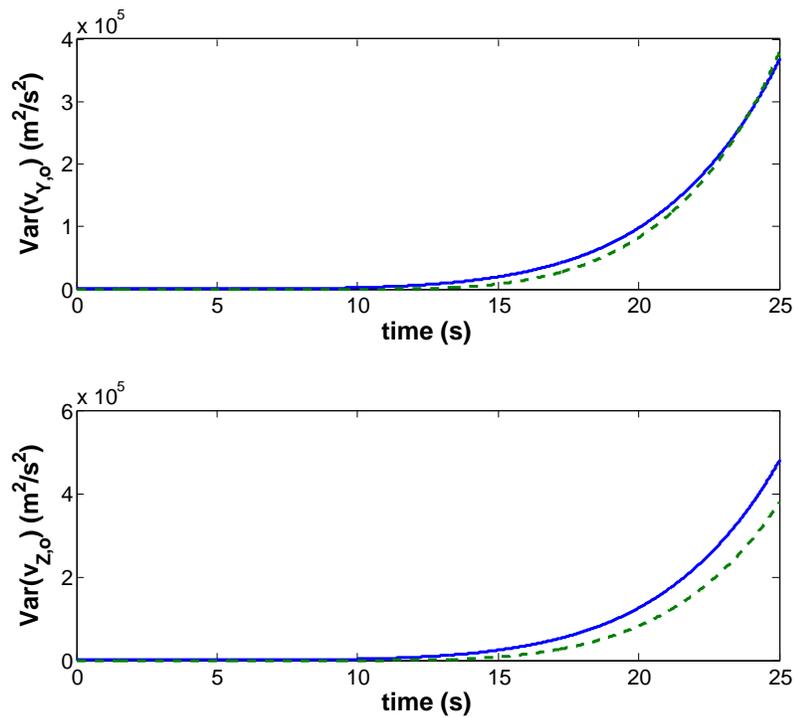}
\caption{Comparison of lateral rocket velocity variance
evolution as estimated using Model I (solid line)
and as computed via Model II (dashed line). The side load period
ends at $ t=10.85 \ \mathrm{s} .$ }
\end{figure}
 
\begin{figure}[!hb]
\centering
\includegraphics[width=4.75in]{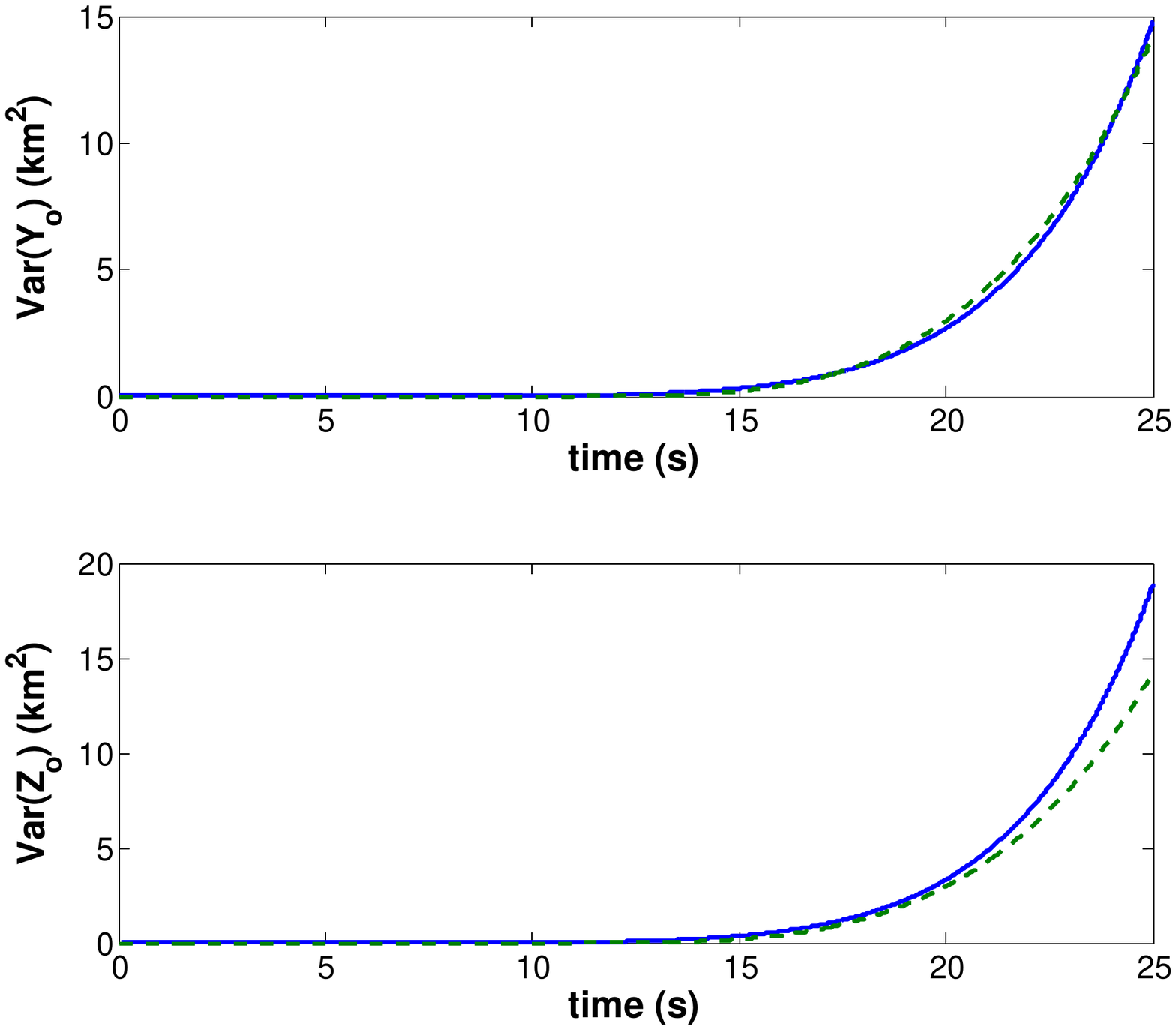}
\caption{Comparison of lateral rocket displacement variance
evolution as estimated using Model I (solid line)
and as computed via Model II (dashed line). The side load period
ends at $ t=10.85 \ \mathrm{s} .$ }
\end{figure}
 
The most important observation, as mentioned
above and as discussed in Appendix A, concerns the post-side-load
emergence of
the thrust force term: as shown by Eqs. 
\eqref{app_eqn_9} and 
\eqref{app_eqn_12}, 
post-side-load lateral velocity and displacement
variances are overwhelmingly determined
by the thrust force component acting in the random
pitch and yaw directions.

Insight into variance growth over the entire flight period
can be gained by interpreting
and scaling individual terms contributing to the velocity
variance in 
\eqref{GrindEQ__82_}
and
\eqref{GrindEQ__83_}. 
Thus, taking the side load period first,
the terms on the right side of 
\eqref{GrindEQ__82_} correspond, respectively,
to velocity \textit{variance-production} via: i) weak 
random side loads
(weak relative to the
thrust, $ F_T ) , $ ii) translational reaction
to weak mass flux damping of random pitch and yaw rates, 
iii) laterally-acting components of the thrust force,
and iv) coupling of effects ii) and iii).
While $ F_T $ is large, angular displacements
during $ 0 \le t \ < T ,$
relative to those appearing during the post-side-load period,
remain small.

During the post-side-load period,
the side load contribution, i), drops out
and a contribution, 
$ < \dot{\eta}^2 (T) > , $
capturing the lumped, integrated effects of ii) through iv),
replaces it. During this period, angular displacements
grow large enough for large lateral thrust force
components to appear, generating, in turn, large lateral
velocities and displacements.

Notice that lateral displacements,
on the order of several kilometers, are observed.
As discussed in \cite{nilabh2010}, scaling readily shows
that experimental and theoretical displacements are consistent
with the magnitudes of computed side loads, thrust forces
and angular displacements.  Note too, that these
displacements represent small fractions
of the rocket's vertical displacement; 
the difference between ideal, zero-side-load three-dimensional rocket
displacements (relative to the launch location) and displacements
observed when side loads are accounted for, remains less than
0.1 \% over the simulated flight period,
$ (0, T_f ] $ \cite{nilabh2010}.
\section{Summary and conclusions}
A set of analytical models are proposed 
which connect stochastic, shock-induced boundary layer separation
in over-expanded nozzles, to random nozzle side loads
and associated rocket response. The broad objectives center
on establishing the consistency of Models I \cite{nilabh2010}
and II, 
and on broadening 
understanding of the physical features connecting boundary layer
separation and rocket response. 

The physical consistency of Model I is established
by demonstrating that the simple model of stochastic
separation line evolution proposed in \cite{nilabh2010}
allows direct \textit{derivation} of the \textit{ad hoc} side load
model assumed in \cite{nilabh2010}, as well as
derivation of observed side load amplitude
and direction probability densities \cite{ref7, ref7a, ref7b}; the demonstration
thus ties the most uncertain elements of Model I, the separation line
and side load models, to experimental observation.
The consistency of the analytical models proposed here,
collectively Model II, is established by showing that
predicted altitude-dependent statistics
of rocket rotational and translational motion are,
in every case examined, consistent with those predicted
by Model I.

From a practical standpoint, the analytical models proposed here
provide a
simple framework for analyzing
stochastic rocket response to side loading, as well
as identifying design strategies that
minimize either side loads, side load-induced torques,
and/or rocket response to these loads and torques.
\renewcommand{\theequation}{A-\arabic{equation}}
\setcounter{equation}{0}
\section*{Appendix A: problem-specific variances}
Detailed variance expressions
for pitch and yaw rates and angles, and lateral translational 
velocities and displacements, 
appropriate to the set of conditions used in numerical experiments,
are given here.
A few non-trivial derivational details are included.

Variances during the side load period, $ 0 \le \ t < T , $
depend on the diffusion coefficient, $ D(t) , $
and damping coefficient, $ A(t) , $ given respectively by Eqs. 
(\ref{GrindEQ__60_}) and (\ref{GrindEQ__57_}). These in turn depend on the 
time dependent position of the shock in the nozzle, $ x_s (t) $ (as given implicitly by
$ R(t) ),$  the time-dependent
pressure difference, $ \Delta P (t) , $ the time-dependent rocket mass,
$ M(t) , $ and moment of inertia, $ I(t) . $ Here, 
these terms are determined via quadratic fits
to data from numerical experiments; under, e.g.,
real flight conditions, these terms could, in principal, be estimated
using computational fluid dynamics simulations, perhaps combined with
experimental measurements.

\subsection*{Pitch and yaw rate variances}
Pitch and yaw rate variances during the side load period, 
given by Eq. \eqref{GrindEQ__65_},
assume the specific form:
\begin{equation}\label{app_eqn_1}
< \omega_{\alpha}^2 (t)> = (L-b)^2 2 \pi \sigma_s^2 \tau_c \epsilon 
\big[ \frac{I_o - |\dot{I}|t}{I_o} \big]^{c_1} 
\int_0^t \frac{R^2 (s') 
\Delta^2 P (s') }{I^2(s')} \big[ \frac{I_o}{I_o- |\dot{I}|s'} \big]^{c_1} ds'
\quad 0 \le t \le T
\end{equation}
where
\begin{equation}\label{app_eqn_2}
c_1= 2 \left[ \frac{\dot{I}+|\dot{M}|(L-b)^2}{|\dot{I}|} \right]
\end{equation}
and where, due to the fixed magnitude of $ \dot{M} , $ 
\begin{eqnarray*}\label{app_eqn_2a}
M(t) = & M_o - |\dot{M}| t  \\
I(t) = & I_o - |\dot{I}| t 
\end{eqnarray*}
Derivation of \eqref{app_eqn_1}
is straightforward,
requiring separate evaluation of the three integrals
in \eqref{GrindEQ__65_}, with, e.g., the equation 
immediately above used for 
the $ I(t) $ term in the diffusivity, $ D(t) $ (equation \eqref{GrindEQ__60_}),
and by noting throughout that $ \dot{M} $ and $ \dot{I} $ are constant.

During the post-side-load period, $ T <  t \le T_f , $ Eq. \eqref{GrindEQ__69_}
assumes the specific form: 
\begin{equation}\label{app_eqn_3}
\left< \omega_{\alpha}^2(t) \right> = \left< \omega_{\alpha}^2(T) \right> 
\left[\frac{I_o - |\dot{I}|t}{I_o-|\dot{I}|T} \right]^{c_1}
\quad T <  t \le T_f
\end{equation}
\subsection*{Pitch and yaw variances}
The general forms of the side-load-period and post-side-load
pitch/yaw variances are given respectively by \eqref{model-D-eqn-7}
and \eqref{model-D-eqn-10}. The former takes the specific form 
\begin{equation}\label{app_eqn_4}
\left<\psi_{\alpha}^2(t) \right> = 2 \int_0^t \left[ \frac{I_o - |\dot{I}|q'}{I_o}
\right]^{c_1/2} \left( \int_0^{q'} F_1(q) 
\left[ \frac{I_o - |\dot{I}|q}{I_o} \right]^{c_1/2} dq \right) dq'
\quad 0 \le t < T
\end{equation}
where
\begin{equation}\label{app_eqn_5}
F_1(q) = 2 \pi \sigma_2^2 \tau_c \epsilon (L-b)^2 
 \int_0^q \frac{R^2(s') \Delta^2 P(s')}{I^2(s')} \left[ \frac{I_o}{I_o -
|\dot{I}|s'} \right]^{c_1} ds'
\end{equation}
Again, this expression follows via straightforward evaluation
and integration of individual terms in \eqref{model-D-eqn-7}.

Pitch/yaw variance during the post-side-load period is most easily
determined by computing individual terms in \eqref{model-D-eqn-10}.
Thus, 
\begin{equation}\label{app_eqn_6}
\left< \omega^2(T) \right> \left( \int_T^t h(s') ds' \right)^2 =
\frac{\left< \omega^2(T) \right> }{(I_o - |\dot{I}| T)^{c_1}}
\frac{1}{|\dot{I}|(1+c_2)} \left[(I_o - |\dot{I}|T)^{c_2+1} -
(I_o - |\dot{I}|t)^{c_2+1} \right]
\end{equation} 
where $ c_2 = c_1/2 =
\left( \dot{I}+|\dot{M}|(L-b)^2 \right)/|\dot{I}| . $
In addition,
\begin{equation}\label{app_eqn_7}
2 \left< \omega_{\alpha}(T) \psi_{\alpha}(T) \right> 
\left( \int_T^t h(s') ds'  \right)  =
2 \left< \omega_{\alpha}(T) \psi_{\alpha}(T) \right> 
\frac{1}{(I_o - |\dot{I}|T)^{c_2}} 
\frac{1}{|\dot{I}|(1+c_2)} \left[(I_o - |\dot{I}|T)^{c_2+1} -
(I_o - |\dot{I}|t)^{c_2+1} \right]
\end{equation}
where the detailed form of 
$ \left< \omega_{\alpha}(T) \psi_{\alpha}(T) \right> $
follows from \eqref{model-D-eqn-13}:
\begin{equation}\label{app_eqn_8}
\left< \omega_{\alpha}(T) \psi_{\alpha}(T) \right> = \int_0^T 
\left[ \frac{I_o - |\dot{I}| q}{I_o} \right]^{c_2}
\left[ \frac{I_o - |\dot{I}| T}{I_o} \right]^{c_2}
\int_0^q D(s') g^2(s') ds' dq
\end{equation}
and where, from \eqref{GrindEQ__60_} and the definition of $ g $
immediately following \eqref{model-D-eqn-13}
\begin{equation}\label{app_eqn_9a}
D(s') g^2(s') = 2 \pi \sigma_s^2 \tau_c \epsilon (L-b)^2 
\left( \frac{R^2(s') \Delta^2 P(s')}{I^2(s')} \right)
\left[ \frac{I_o}{I_o - |\dot{I}|s'} \right]^{c_1}
\end{equation}

The final term in \eqref{model-D-eqn-10}, $ < \psi_{\alpha}^2(T) > , $
follows by setting $ t =T $ in 
\eqref{app_eqn_4}.
\subsection*{Lateral translational velocity and displacement variances}
Full expressions for lateral translational velocity variances 
during and subsequent to the side load period are given respectively by
\eqref{GrindEQ__82_} and \eqref{GrindEQ__83_}.  Corresponding
displacement variances are given by 
\eqref{GrindEQ__90_} and \eqref{GrindEQ__91_}.  
Considering the entire flight period, $ 0 \le t \le T_f , $
we find via scaling and numerical experiments
that in the present model, velocity and displacement variances
remain small during the side load period, but grow explosively
during the post-side-load period. 

Due to the dominance of post-side-load variances,
for simplicity, and as mentioned, 
when computing theoretical velocity and displacement
variances, we simply compute $ < \dot{\eta}^2 (t) > $
and $ < \eta^2(t) > $ during the post-side-load period, using
only the terms involving the thrust force, $ F_T . $
It is readily shown that these terms are, at minimum, two orders of magnitude larger
than all other, neglected terms.

Thus, the post-side-load velocity variance assumes the form:
\begin{equation}\label{app_eqn_9}
\left< \dot{\eta}^2 (t) \right> = \int_T^t \int_T^t
\frac{F_T(s') F_T(p')}{M(s')M(p')} \left< \psi_{\alpha \pm}(s') 
\psi_{\alpha \pm} (p') \right> ds' dp'
\end{equation}
The time correlation function, in turn, is given as the weighted sum
of three separate variances:
\begin{equation}\label{app_eqn_10}
\left< \psi_{\alpha \pm} (s') \psi_{\alpha \pm} (p') \right> =
k(s') k(p') \left< \omega_{\alpha \pm}^2 (T) \right> +
\left< \psi_{\alpha \pm}^2 (T) \right> +
\left[k(s') + k(p') \right] \left< \omega_{\alpha \pm}(T) \psi_{\alpha \pm} (T) \right>
\end{equation}
where 
\begin{equation}\label{app_eqn_11}
k(q) = \frac{1}{(I_o - |\dot{I}|T)^{c_2}} 
\frac{1}{|\dot{I}|(1+c_2)} \left[(I_o - |\dot{I}|T)^{c_2+1} -
(I_o - |\dot{I}|q)^{c_2+1} \right]
\end{equation}
The term $ \left< \omega_{\alpha \pm}^2 (T) \right> $
follows by using $ t = T $ in \eqref{app_eqn_1}, 
$ < \psi_{\alpha}^2(T) > , $
follows by doing the same in 
\eqref{app_eqn_4}, and 
$ \left< \omega_{\alpha \pm}(T) \psi_{\alpha \pm} (T) \right> $
is given by \eqref{app_eqn_8}.

Finally, lateral displacement variance during the post-side-load period
is given by
\begin{equation}\label{app_eqn_12}
\left< {\eta}^2 (t) \right> = \int_T^t \int_T^t
\left< \tilde{F}(s') \tilde{F}(p') \right> ds' dp'
\end{equation}
where
\begin{equation}\label{app_eqn_13}
\left< \tilde{F}(q) \tilde{F}(q') \right> = 
\int_T^{q'} \int_T^q
\frac{F_T(s') F_T(s)}{M(s')M(s)} \left< \psi_{\alpha \pm}(s') 
\psi_{\alpha \pm} (s) \right> ds' ds
\end{equation}
\renewcommand{\theequation}{B-\arabic{equation}}
\setcounter{equation}{0}
\section*{Appendix B: Random wind effects}
When considering the dynamic effects of wind loading, it is useful to focus 
on the ratio of the characteristic time scale associated with say a turbulent 
cross-wind, $ \tau_w = l_w / v_w^Ë, $ relative to the rocket dynamics time 
scale, $ \tau_R = L_R/V_R : $ 
\begin{equation}\label{windratio}
\frac{\tau_w}{\tau_R} = \frac{l_w}{L_R} \frac{V_R}{v_w^Ë}
\end{equation}
where $ l_w $ and $ v_w^Ë $ are the integral length and velocity scales for 
the cross-wind, $ L_R $ is the rocket length, and $ V_R  $ is the 
characteristic rocket speed.  Given reasonable, readily estimated velocity 
scales, then wind loading produces non-negligible pitch and yaw responses 
only under conditions where the length scale ratio produces time scale ratios 
of order one or less.  Under these circumstances, the turbulent wind loads 
act fast enough to produce non-negligible pitch and yaw moments.  In 
contrast, when the turbulence time scale is much larger than the rocket 
dynamics time scale, then lateral wind turbulence merely produces random 
lateral displacements, with minimal pitch and yaw (where the random 
translational 
response presumably has zero mean relative to the mean displacements and 
velocities produced by the mean cross-wind).  

Based, e.g., on a large number of altitude-dependent wind measurements (taken 
under nominally normal daytime conditions at Cape Kennedy) \cite{nasa-wind-
load-report},  $ v_w^Ë = O(10 \ \mathrm{m/s} ), $ while $ l_w = O(10^3 \ 
\mathrm{m}) ,$ where the latter corresponds to the approximate correlation 
length scale.  Thus, since $ V_R = O(10^3 \ \mathrm{m/s}) $ and $ L_R = 10 \ 
\mathrm{m} , $ $ \tau_w/\tau_R = O(10^4) .$ While prevailing winds thus have 
negligible effect on representative rocket rotational dynamics, the wind 
nevertheless produces a random, translational response.  Indeed, a quick 
estimate shows that the lateral wind load $(=O(\rho_a \bar{V}_w^2 A_R), $ 
where $ A_R $ is the rocket lateral area) is of the same order as the 
characteristic side load (estimated below); however, since the side load time 
scale is fast, specifically, faster than the rocket time scale, $ \tau_R , $ 
then the 
slow-time scale dynamics produced by wind can be neglected when studying the 
response produced by side loads.  [When computing the 
combined effect of random wind and side loading, due to the separation in 
response time scales, the response to the former can be simply superposed on 
that produced by the latter.]

\end{document}